\documentclass[twocolumn]{aastex631}

\usepackage{tabularx}
\usepackage{hyperref}
\usepackage{tabularx}

\begin{document}

\title{Analysis of Newly Catalogued Open Star Cluster UPK~220 with Gaia DR3 and TESS: Discovering Member Variable Stars}

\correspondingauthor{İnci Akkaya Oralhan}
\email{iakkaya@erciyes.edu.tr}

\author[0000-0003-1787-7418]{İncİ Akkaya Oralhan}
\affiliation{Department of Astronomy and Space Sciences, \\
Science Faculty, Erciyes University, \\
38030 Melikgazi, Kayseri, Türkiye}
\affiliation{Astronomy and Space Sciences Observatory and Research Center (Uzaybimer), \\
Erciyes University, 38281, Talas, Kayseri, Türkiye}


\author[0000-0001-9198-2289]{Cenk Kayhan}
\affiliation{Scientific Research Projects Coordination Unit, \\
Kayseri University, \\
38280, Talas, Kayseri, Türkiye}

\author[0000-0002-3983-6628]{Özgün Arslan}
\affiliation{Department of Astronomy and Space Sciences, \\
Science Faculty, Erciyes University, \\
38030 Melikgazi, Kayseri, Türkiye}
\affiliation{Astronomy and Space Sciences Observatory and Research Center (Uzaybimer), \\
Erciyes University, 38281, Talas, Kayseri, Türkiye}



\begin{abstract}
Studies on star clusters with the same age and initial chemical composition have gained momentum in recent years with the use of \textit{Gaia}. In addition, the discovery of new clusters with 
Gaia has increased the number of open clusters to be examined. Many of these discovered 
sources are intermediate-age open clusters and have not been analyzed 
in detail yet. 
In this study, we focused on newly cataloged open cluster UPK~220. 
The fundamental parameters (distance, age, metallicity and reddening) of UPK~220 were determined by analysing the variable stars within the cluster, while simultaneously constraining the parameters of the variable stars using these cluster parameters. To achieve this, we combined GaiaDR3 and TESS photometric observations. Using GaiaDR3, we derive fundamental parameters of UPK~220 through membership analyses, and with TESS, we discovered eight member variable stars.
We also extracted the atmospheric 
parameters ($logg$, $[Fe/H]$ and $T_{\rm eff}$) for the variable stars using SED, GSP-Phot and GSP-Spec, and MESA models.

\end{abstract}


\section{Introduction} 
\label{sec:intro}

Open clusters (OCs) are gravitationally bound  stellar populations, which show characteristic number density profiles of member stars with respect to field stars \citep{1966AJ.....71...64K, krumholz2019star,sariya2023gaia}. According to their ages, OCs can be classified into three main categories: 
young ($ < 10^{6} $ yr), intermediate-age 
($ 10^{6} - 10^{7}$ yr) and old ($ > 10^{7}$ yr) \citep{friel1995old,im2023photometry}.

OCs may involve a few hundred to thousand member stars depending on the initial physical conditions of the molecular cloud regions where they formed and their galactic location within the Galactic disc.
Therefore, OCs are paradigmatic coeval systems to investigate the 
formation of stars and stellar evolution but also critically important to understand the chemical and dynamical structure within the Galactic disc \citep{2007ApJ...669.1167B,kim2017bvi,cantat2018gaia,2020AA.640A..1C}.

The predicted total number of galactic OCs is near of $10^{5}$ \citep{piskunov2006revisiting}. For this reason, many extensive studies have been carried out to identify candidate OCs, as well as to comprise advance catalogs of OCs based on multiple photometric surveys \citep{mermilliod1981comparative,dias2002new,cutri2003vizier,kharchenko2005astrophysical, kharchenko2013global}. 
However, the number of known OCs within available catalogs produced before the Gaia Mission \citep{ prusti2016gaia} is substantially restricted due to photometric limits, field star contamination, spatial resolution, and dust extinction \citep{he2022new, chi2023identifying}. 

The census of discovered OCs dramatically increased with 
the Gaia Data Release 2 (hereafter GaiaDR2) \citep{ cantat2018gaia, soubiran2018open, liu2019catalog, castro2020hunting,monteiro2020fundamental}.  
Considering certain astrometric and photometric data such as equatorial coordinates (${\alpha,  \delta}$), proper motions ($\mu_{\rm \alpha},\mu_{\rm \delta}$), trigonometric parallaxes ($\varpi$), and magnitudes (the white-light G, the blue $G_{BP}$, and the red $G_{RP}$) with the accompanied errors, the effective separation between member stars and field stars across an OC region can be achieved \citep{uribe2002membership,gao2018memberships}.
Gaia Data Release 3 \citep[hereafter GaiaDR3]{2023A&A...674A...1G}, an updated version of GaiaDR2, provides quite a suitable and homogeneous data set to investigate and promote the fundamental parameters of Galactic OCs \citep{he2022unveiling}.
The database comprises nearly 1.8 billion celestial objects with magnitudes up to 19th in the G band, presenting fairly homogeneous and low photometric uncertainties \citep{negueruela2023gaia}. Therefore, GaiaDR3 gives us an ideal data set to study high-precision photometric analyses of OCs and
the parameters of a cluster can be determined with greater precision  
\citep{lada2003embedded,yadav2011optical, ahumada2013ngc,pietrzynski2019distance,castro2020hunting}.

The mass distribution of OC members result in the stars undergoing disparate evolutionary stages despite their identical age \citep{bonatto2005detailed,bonatto2006mass}. 
In addition to the evolutionary stages observed, 
some members have been identified as variable stars.
These variable member stars are classified mainly into two subtypes
that are based on intrinsic and extrinsic variations \citep{xin2002searching, mowlavi2013stellar, chehlaeh2018photometric,  dar2018search, jiang2020case, kharchenko2022evolutionary, wang2022searching, zhuo2021variable}. 

Studying variable stars in an OC will bring constraints to the determination of both parameters of the cluster and variable member stars. In this way, the physical parameters of the cluster, such as distance ($d$), age ($t$), metallicity ($[Fe/H]$), and interstellar reddening 
($E(G_{BP}-G_{RP})$) that is determined from GaiaDR3 data, are used as input parameters in constructing the models of the variable stars.
Thus, the model parameters of these discovered variable stars will be determined more precisely \citep{durgapal2020search, shen2021search}.
  
In this study, the variable member stars detected by Transiting Exoplanet Survey Satellite \citep[hereafter TESS,][]{Ricker2014,Ricker2015} photometric observations. 
TESS is a crucial telescope for the detection of variable stars. 
TESS possesses comprehensive sky scanning capabilities and high temporal resolution, with two distinct cadences: short (120 sec) and long (30 min) \citep{sun2022search}.  

The majority of OC studies focus on the fundamental parameters of the
cluster, with relatively little emphasis on individual members or the influence of those overall cluster properties.
However, there are several studies in which both cluster membership analyses are conducted and 
the parameters of the member variable stars are compared with the cluster fundamental parameters
(e.g. \citet{2023ApJ...946L..10B, 2023A&A...677A.154F, 2023MNRAS.520.1092L, 2024A&A...681A..13F}). 
It is therefore important to perform studies that make such comparisons while integrating data from different databases. These efforts enhance the sensitivity of parameter determinations for both the cluster and its members. Moreover, members of OCs may include photometric or spectroscopic variable stars. In such cases,
the variable stars within OC members may serve to reduce  
uncertainties in the fundamental parameters of the cluster, given the fact that they share the same age and initial chemical composition. 

In this study, the most up-to-date database, GaiaDR3, is employed for the high-precision analysis of cluster membership. Additionally, TESS observations are utilized to identify photometric variable stars in the studied cluster UPK~220.
UPK~220 is a newly discovered intermediate-age 
OC included in the Ulsan Pusan Korea (UPK) star cluster catalog. 
The UPK catalog exemplifies the efficiency of GaiaDR2 in detecting new OCs within the Galactic disc \citep{ sim2019207}. UPK~220 is located in the second Galactic quadrant at a distance of 967 pc, with an estimated age of 560 Myr \citep{2019JKAS...52..145S}. Its central equatorial coordinates {\rm $\alpha_{J2000} = 23^h 23^m 46.1^s$, $\rm \delta_{J2000} = +66\degr 30\arcmin 18\arcsec$} are recently estimated by \citet{tarricq20213d}.
Its astrophysical parameters have not been comprehensively revised, and discrepancies in its estimated $t$, $d$, and $[Fe/H]$ have been reported in the literature \citep{2020AA.640A..1C,cavallo2023parameter,2024AA...689A..18A}. 

This paper is organized as follows: the membership selection and member variability are presented in Section~\ref{sec:data}. The determination of 
the cluster parameters and the analysis results of variable
stars are presented in Sections~\ref{sec:cluspar} and \ref{sec:resvar}.
Finally, Discussion and Conclusion are 
summarized in Sections~\ref{sec:diss} and ~\ref{sec:conc}.

\section{Data and Method} 
\label{sec:data}

This section outlines the methodology used for the membership analysis of the stars
within the cluster region using GaiaDR3 data. It also details the procedures employed to determine whether the members exhibit variability and to classify their variable types based on photometric variations observed by TESS data.

\subsection{Membership Selection}
\label{sec:pyupmask}

Analyzes of the OCs is difficult by the presence of contaminating 
foreground and
background stars in the projected field of view. Therefore, it is essential to distinguish cluster members from field stars using a decontamination algorithm. In this study, we employ the {\texttt{pyUPMASK}} algorithm, an enhanced version of the original {\texttt{UPMASK}} algorithm \citep{2021A&A...650A.109P}. Since UPK~220 spans a relatively wide field of view ($\sim$0.5$^\circ$) on the sky, the parameters $\mu_{\alpha}$ and $\mu_{\delta}$, $\varpi$ and G, $\rm G_{BP}$ and $\rm G_{RP}$ from 
GaiaDR3, along with their uncertainties, were extracted for stars within a 30-\textit{arcmin} radius to apply {\texttt{pyUPMASK}}. 

\begin{figure*}
	\includegraphics[width=13cm, angle=270]{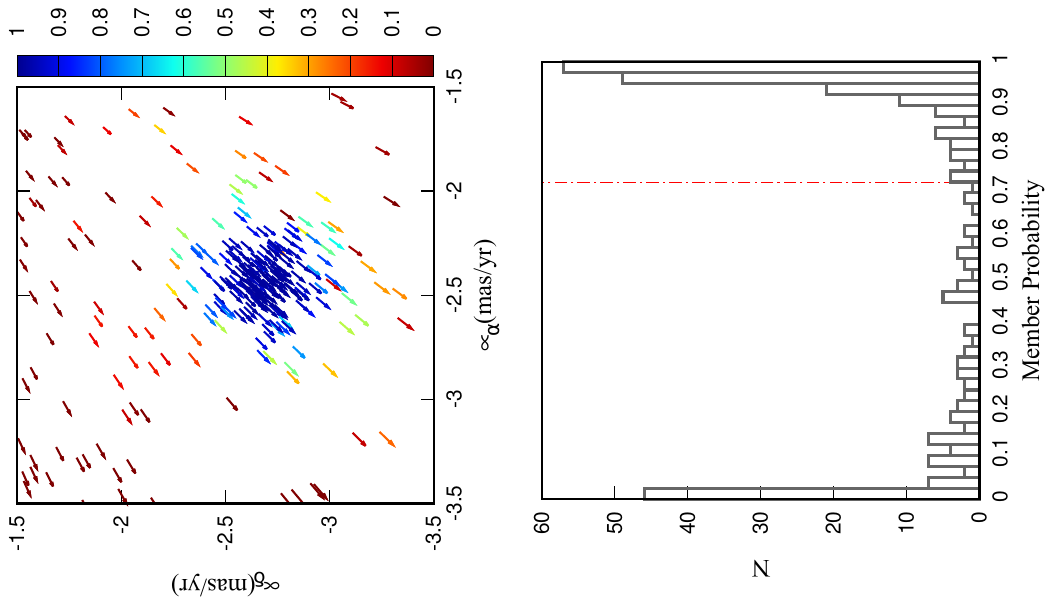}
    \caption{\textit{Upper panel:} The proper motion vector diagram of UPK~220. The right colour bar represents the membership probabilities. \textit{Lower panel:} The distributions of membership probabilities according to \texttt{pyUPMASK}. The vertical red dashed line shows the selected probability limit of 0.725.}
    \label{fig:ppm}
\end{figure*}

In the proper motion vector diagram of UPK~220 shown in Figure~\ref{fig:ppm} (\textit{upper panel}), we restricted the sample to stars with parallaxes in the range $0.95 \leq \varpi<1.1$ \textit{mas} based on the distance estimates from \citet{2019JKAS...52..145S}. This selection effectively removed many foreground and background stars. The histogram of membership probabilities obtained from {\texttt{pyUPMASK}} is displayed in Figure~\ref{fig:ppm} (\textit{lower panel}) for 284 stars. Based on these probabilities, we classified stars with a membership probability $P \geq 0.725$ as cluster members, yielding 163 members. The red dashed line in the histogram represents the threshold at which 
the star counts begin to increase, thereby distinguishing the member stars from field stars. 
From the member stars, the median parallax and proper motion values for the cluster members were determined to be $\varpi = 1.03 \pm 0.03$ $mas$ and ($\mu_{\rm \alpha}, \mu_{\rm \delta}) = (-2.41 \pm 0.11, -2.64 \pm 0.12$) \textit{mas/yr}, respectively.
For comparison, \citet{2019JKAS...52..145S} reported proper motion values of ($\mu_{\rm \alpha}, \mu_{\rm \delta}$)=
 ($-2.41 \pm 0.13, -2.59 \pm 0.12$) \textit{mas/yr}. 

\begin{figure*}
\includegraphics[scale=0.5,width=9 cm, angle=0]{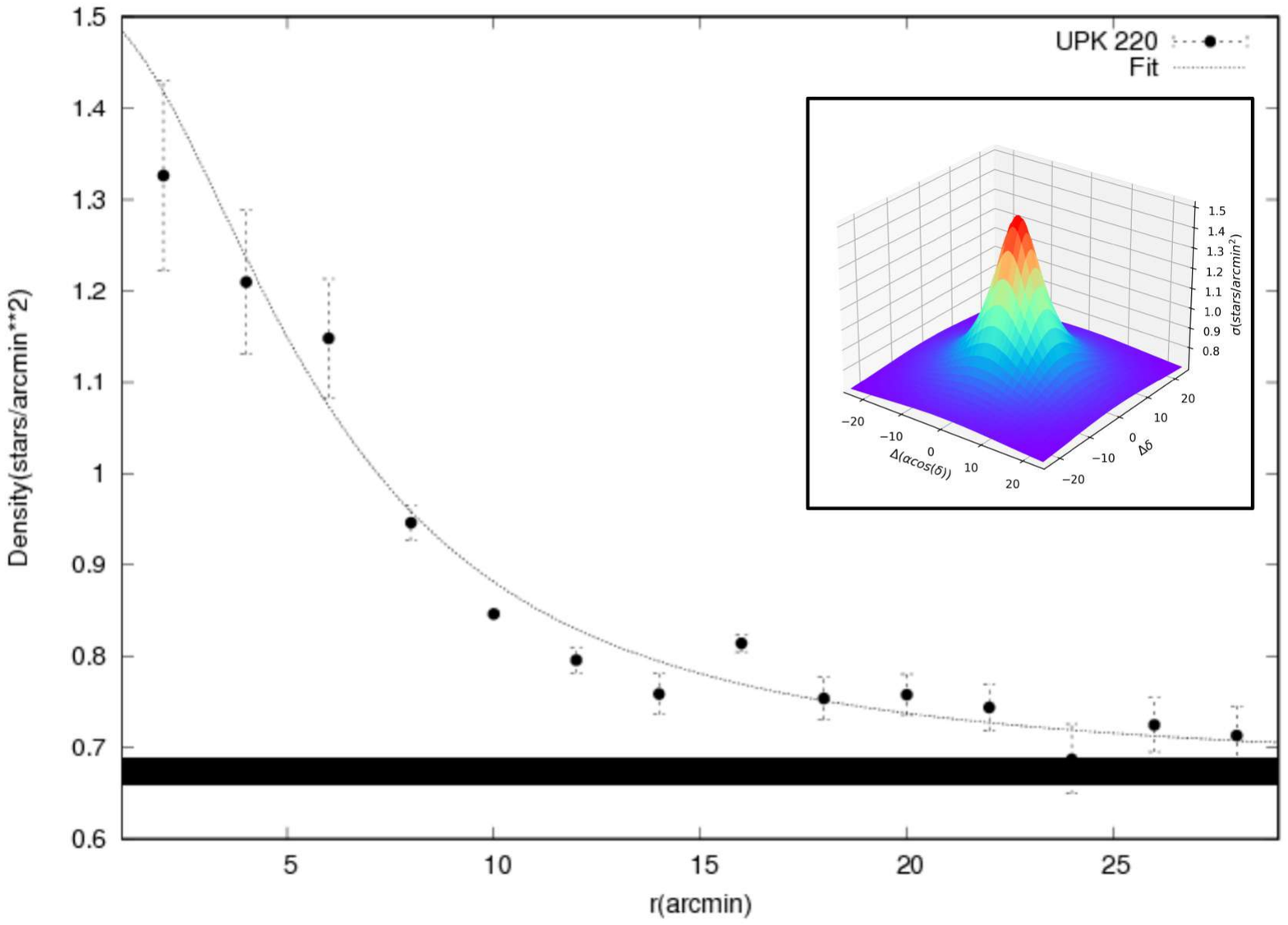}
    \caption{The radial density profile of UPK~220. The dotted curved line shows the fitting of King's profile. The horizontal black bar denotes the stellar background level measured in the comparison field. \textit{Inserted plot on top}: The three-dimensional stellar surface density of UPK~220 accomplished by astrometric and photometric field star decontamination procedures on the cluster region.}
    \label{fig:king}
\end{figure*}

To determine the size of UPK~220, its stellar radial density profile (RDP) was constructed  using GaiaDR3 data 
by counting stars within concentric rings of increasing width centered on the cluster as see in Figure~\ref{fig:king}.
From a King profile fit \citep{1966AJ.....71...64K}, we derived the central stellar density ($\sigma_{\rm {ok}}$), core radius (R$_{c}$) and residual background density ($\sigma_{\rm bg}$) for UPK~220. 
The obtained parameters are $\sigma_{\rm_{ok}}= 0.835 \pm 0.121 $ $stars/{\rm \textit{arcmin}^{2}}$, R$_{c}= 5.751 \pm 
1.079$ \textit{arcmin}, and $\sigma_{\rm bg} = 0.674 \pm 0.023$ $stars/{\rm \textit{arcmin}^{2}}$. 
According to the RDP, we adopt 
the cluster limit radius $R_{\rm lim}= 24.0$ \textit{arcmin} as the outer boundary. Considering King's profile fitting, we accept 148 stars within UPK~220 limit radius as the probable cluster members. 

\subsection{Members Variability}

\begin{figure}
    \includegraphics[width=7 cm, angle=270]{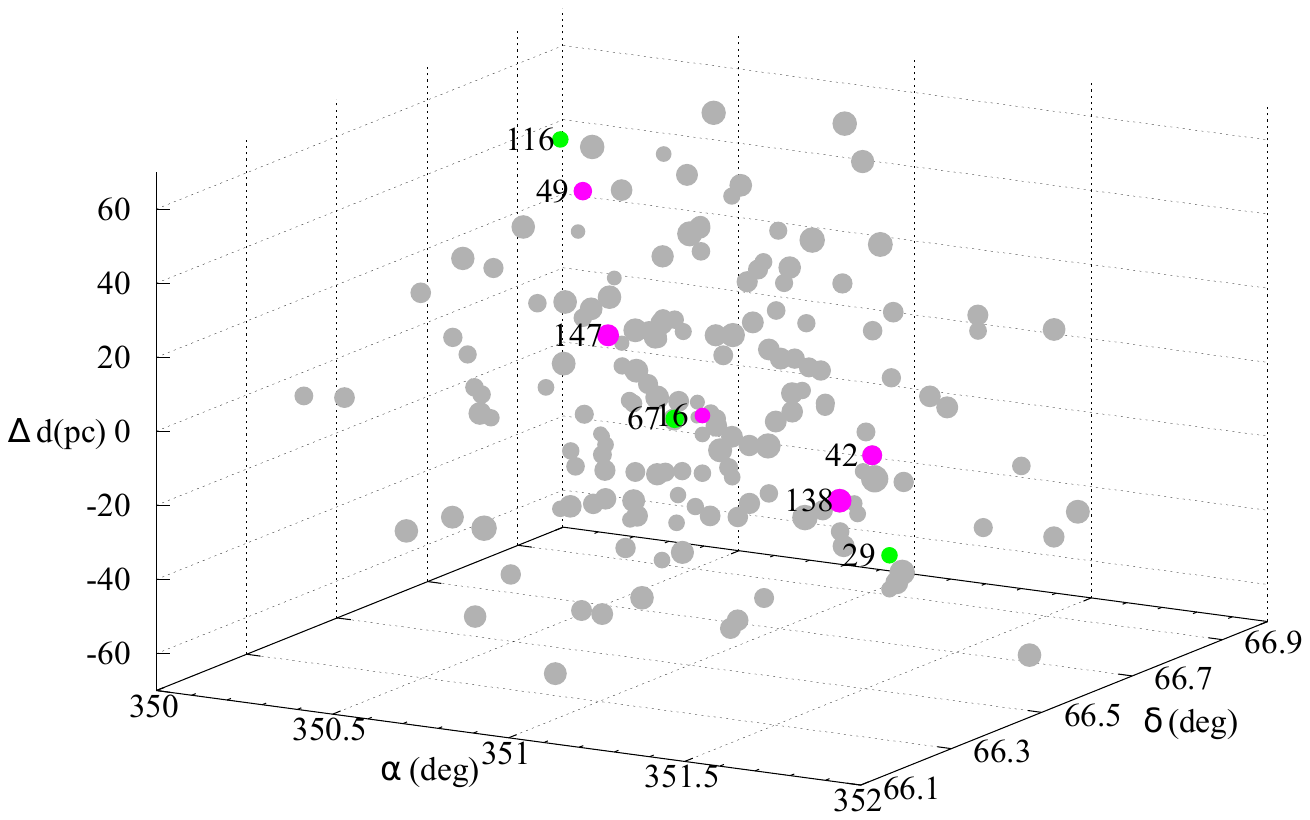}
    \caption{Three dimensional distribution of the cluster members with eight member variable stars. The dot size of the stars corresponds to $G_{\rm mag}$.}
    \label{fig:3D}
\end{figure}

UPK~220 is observed by TESS in five different sectors 
with 30-min (sectors 17, 18 and 24) and 2-min (sectors 57 and 58) cadences. 
The data that is produced by Science Processing Operations 
Center \citep[SPOC,][]{2016SPIE.9913E..3EJ} is downloaded from 
Barbara A. Mikulski Archive for Space Telescopes (MAST)
database{\footnote{\url{https://mast.stsci.edu/portal/Mashup/Clients/Mast/Portal.html}}}.
TESS total observed time of member stars UPK~220 is 124.29 days. 
However, there are 
two significant gaps in the observation timeline: one between sectors 18-24, and another between sectors 24-57{\footnote{Details can be 
found in the TESS Data Release Notes (DRN). For Sectors 17, 18, 24, 57 and 58 (DRN 24, 
25, 35, 82 and 83, respectively), they are available at 
{\url{https://archive.stsci.edu/tess/tess\_drn.html}.}}}. 
We analyzed the photometric data of UPK~220 using Full-Frame Image (FFI) obtained by TESS. We used \texttt{lightkurve} code \citep{Lightkurve} to obtain light curves of UPK~220 members
from FFIs. To minimize to contamination from nearby stars, we employed TESScut~\citep{2019ascl.soft05007B} to extract cutouts of the FFIs for each target. Since the resolution of TESS is limited \citep[1 TESS pixel = $\sim{21}$ \textit{arcsec},][]{Ricker2015}, we used varying sizes of target and background apertures based on the brightness of the targets. 
After removing background and other light contributions, we produced light curves of each members. 
From these light curves, we identified eight member stars exhibiting variability, listed in Table~\ref{tab:variable_table}. The star IDs correspond to their cluster membership designations. Among these, three stars (IDs 29, 67 and 116) are classified as eclipsing binaries, two stars (IDs 49 and 138) 
are pulsating, two stars (IDs 16 and 42) are magnetic active, and one star (ID 147) is a rotating variable. Figure~\ref{fig:3D} further illustrates the positioning of these stars within the boundaries of the cluster.

We computed the Fourier spectrum and their periodogram for 
the variable stars within a period range of 0.01 to 27 days, 
corresponding to the observation duration of the TESS sectors. 
For each variable star, we calculated the signal-to-noise ratio (SNR) of 
the highest amplitude peak in each frequency spectrum. Periodic variability was considered valid for frequency peaks with an SNR of four or higher, as recommended by \citet{1993A&A...271..482B}. 
According to their variability, we classified based on the shape of their light curves and their period range.

\subsubsection{Eclipsing binary}

For eclipsing binaries, we modeled their light curves 
using v43 version of the {\texttt{JKTEBOP}} code 
\citep{2004MNRAS.351.1277S}. The code has fast modeling feature by 
simulating eclipsing binaries. It has also detailed error analysis including 
Monte Carlo error analysis algorithms. 
However, intrinsic variations such as 
pulsation or stellar spots cannot be included in the light curve models of 
calculations in the code. 

In this study, we modeled light 
curves of eclipsing binaries IDs 29, 67 and 116, 
which were obtained from TESS observations in sectors 17, 18, 24, 57 and 58. 
We also determined 1-$\sigma$ uncertainties using Monte Carlo and residual-
permutation simulations. We noticed that there are few data points around 
the ingress and egress phases in sectors 17, 18, and 24 due to 
the long cadence mode with a 30-min sampling rate. 
Since this situation affects the precision
of the fractional radii derived from the best-fitting model and could 
lead to erroneous results. Therefore, the light curves obtained from 
sectors 57 and 58 observations of TESS (short cadence mode with 120-sec sampling rate) 
that have more data points
around the egress and ingress phases are decisive in the models. 

\begin{figure}
		\includegraphics[width=\columnwidth]{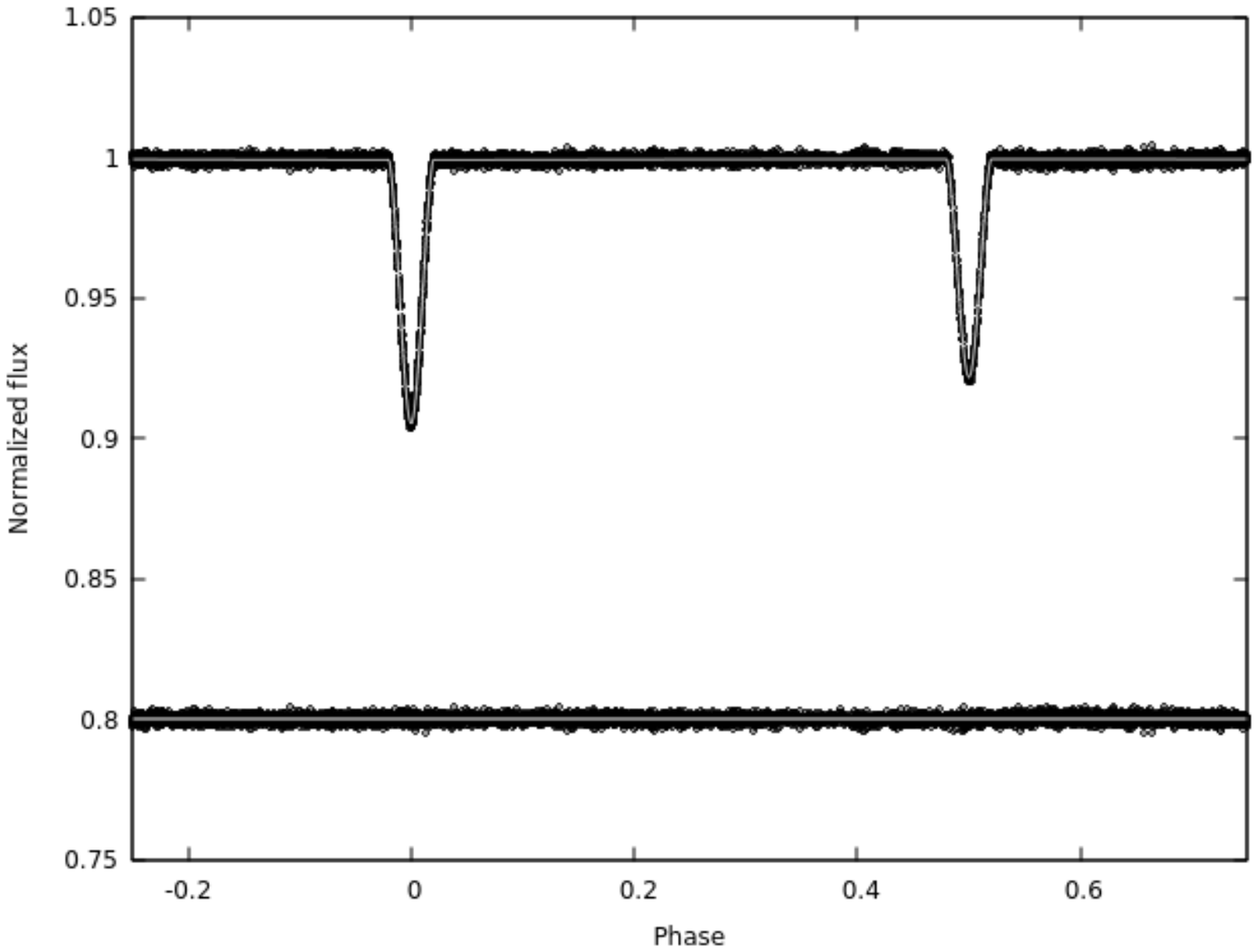}
    \caption{The best fit model (solid line) and residuals (at 0.8) that is obtained by {\texttt{jktebop}} 
and observed TESS light curve of ID 116 at sectors 57 and 58. From the 
    observed primary minima, 
    the ephemeris (E) and phases determined as 
    $T_{\rm primin}=(2459893.26611 \pm 0.00015)+(7.56662 \pm 0.00013)E$.}
    \label{fig:modellc2}
\end{figure}

To prepare the data for light curve modeling, data  
is converted the TESS flux to ${T_{\rm mag}}$ using Pogson 
equation. The initial 
orbital period ($P_{\rm orb}$) and time of periastron of the 
primary eclipse ($T_p$), orbital inclination ($i$),
periastron longitude omega ($w$),
fractional radius of stars ($r_A$ and $r_B$), orbital 
eccentricity ($e$), light ratio (${l_B}/{l_A}$)
are adjusted. In order to 
estimate the linear and non-linear limb darkening 
coefficients, we interpolate 
the values computed for the TESS bandpass in 
\citet{2017A&A...600A..30C}.
As an example, our best-fitting light curve model, 
phase-folded TESS observation data, and  
residuals of ID~116 are shown in Figure~\ref{fig:modellc2}.
To obtain formal uncertainties for the light curve model parameters, 
Monte Carlo simulations are applied using the {\texttt{JKTEBOP}} {\sc TASK8} feature. 
The best-fit solutions for TESS light curves of 
IDs 29, 67 and 116 using {\texttt{jktebop}} code 
and determined the 1-{$\sigma$ } uncertainties using Monte Carlo and residual-permutation 
simulations are represented in Table~\ref{tab:tablelcs}. 
For the light curve models of ID 29 and 116, 
we assume circular orbit. However, $ecc*cos(omega)=-0.0041462148 \pm 0.0000078612$, $ecc*sin(omega)=-0.5891148379 \pm 0.0017385892$ and orbital eccentricity ($e=0.5891294283$), periastron longitude omega ($w=269.5967565681 \deg$) for ID 67.

In addition to the photometric changes observed in the member binary stars resulting from
eclipsing, we also noted additional intrinsic or extrinsic brightness variations in the residuals of the fits for these stars. 
These variations are not included in the light curve models and are treated separately (see in below).

\subsubsection{Other variable stars}

We perform periodogram analysis of the variations for the other 
discovered variable stars in UPK~220.
We classify their variability type based on amplitude, period 
and behavior brightness changes in their
light curves. A detailed analysis of each star and the results are presented in Section~\ref{sec:othervar}.

\subsection{Stellar Atmospheric Parameters and SED Analysis of the Variable Members}

Stellar atmospheric parameters are derived from photometry for eight member variable stars.
For seven of these stars (IDs 29, 42, 49, 67, 116, 138 and 147), 
the surface gravity ($logg$), 
metallicity ($[Fe/H]$) and the effective temperature ($T_{\rm eff}$) are extracted 
using General Stellar Parameterize from Photometry (GSP-Phot) 
and Spectroscopy (GSP-Spec) methods
which include four different algorithms \citep{2023A&A...674A..27A}. 
These methods fit the low-resolution 
$G_{BP}/G_{RP}$ spectra, G and parallax values of the stars.

For 
spectral energy distribution (SED) fitting, we use the open-source python package 
\texttt{ARIADNE} \citep{10.1093/mnras/stac956}.
The package contains six different stellar atmosphere model grids 
and multiwavelength photometry to estimate $T_{\rm eff}$, $[Fe/H]$ 
and $logg$ of the stars.
The SED fitting performed using \textit{Bayesian Model Averaging} to obtain the best-fitted
model parameters. To fit each SED of the variable member stars, 
we collected \textit{2MASS} (J, H and $K_s$), \textit{GaiaDR3} (G, ${G_{\rm BP}}$, and ${G_{\rm RP}}$), \textit{Johnson} (U, B, V, R, and
I), \textit{PS1} (g,i, r, y, and z), \textit{SDSS} (u, g, r, and i), \textit{TESS} (Tmag), 
and \textit{WISE} (W1, W2, W3 and W4) photometric data. It was not possible to 
obtain all the photometric data for each star.
The data used in the SED fitting  are listed in Appendix  
Table~\ref{tab:variable_sedphoto}.

The atmospheric parameters of the stars derived from Gaia data and SED fitting are 
listed in Table~\ref{tab:magbin} for binaries and 
Table~\ref{tab:specphoto_variable_table} for single stars.

\section{Determination of Fundamental Cluster Parameters}
\label{sec:cluspar}

To estimate the fundamental parameters of UPK~220, we focus 
on the member variable stars. 
We construct models of the stars using the Modules for 
Experiments in Stellar Astrophysics (MESA - 
\citet{2011ApJS..192....3P, 2015ApJS..220...15P, 2018ApJS..234...34P, 2019ApJS..243...10P}) 
stellar evolution code. 
Subsequently, the parameters derived from the variable stars 
have been used 
as model input parameters when constructing 
cluster isochrones in \textit{MESA Isochrones and Stellar Tracks}
 \citep[MIST]{2016ApJS..222....8D, 2016ApJ...823..102C}. 
In constructing stellar interior models with \textit{MESA}, 
we employ standard mixing length
theory \citep{1958ZA.....46..108B} for convection treatment and utilise the Herwig approximation 
\citep{2000AA...360..952H} for convective overshooting. OPAL opacity tables 
are taken from \citet{1993ApJ...412..752I, 1996ApJ...464..943I} and nuclear reaction rates 
are used from \citet{1999NuPhA.656....3A} with updated by \citet{2002ApJ...567..643K} and 
\citet{2010ApJS..189..240C} in the models. 
We employ elemental diffusion from the \textit{MESA} default option 
for stellar models that have a mass below 1.2 $M_{\rm \odot}$. 
We construct interior models using \texttt{MESAstar} and \texttt{MESAbinary} packages 
to evolve single and binary stars, respectively. 

\subsection{Metallicity Determination}
\label{metal}

The stars within a specific cluster originate from a vast molecular cloud. Assuming the initial substances within the cloud are thoroughly blended, we can infer that all members of the cluster possess identical metallicity. 
This knowledge is essential for studying the local and global properties of galaxies. However, dealing with metallicity is not straightforward. 
While isochrone fitting in the color-magnitude diagram (CMD) can provide fairly precise results \citep{2021MNRAS.504..356D}, it is not very sensitive to variations in metallicity.  

In recent years, both individual studies and large-scale spectroscopic surveys, such as the one discussed by \citet{2022MNRAS.509..421N}, have significantly advanced our understanding of the Galactic metallicity distribution based on OCs. 
Despite the increased availability of physical parameters for numerous clusters from the Gaia database, metallicity remains a source of uncertainty. Even known metallicity values for clusters often stem from spectral data of only one or a few stars. 

In light of this, our study aims to enhance parameter estimation by leveraging distinctive features of specific stars within the cluster, utilizing TESS data from member stars. 

We employed an automated tool called \texttt{Metalcode}{\footnote{\url{https://github.com/mpiecka/metalcode}}}  
for the determination of parameters of clusters, such as reddening, age, and metallicity, applying iterative processes derived from methods developed by \citet{poehnl2010statistical}.
In Metalcode, astrophysical parameters of UPK~220 are derived from isochrone grids
by eliminating distance and reddening as free parameters. This allows the tool to focus on deriving the metallicity of the cluster with high precision.
The initial estimations are iteratively refined by comparing observed data in the normalised temperature-luminosity diagram that is driven by the isochrone grids.
Also, \texttt{Metalcode} is capable of computing the three fundamental parameters of the cluster: $E(G_{BP}-G_{RP})$, $t$, and $[Fe/H]$, performed for Johnson $B$ and $V$, 2MASS $J$ and $K_{s}$, and Gaia $\rm G$, $\rm G_{BP}$, and $\rm G_{RP}$ photometric systems \citep{2023OAP....36...77P}. The process involves an automated $\chi^{2}$ minimization technique that matches magnitudes and colors of the members with theoretical isochrones. The code works for the different metal abundance (\textit{Z}) and age (\textit{log(t)}) values, starting from $Z=0.001$ up to $Z=0.040$, with a step as $\Delta$$Z=0.001$, from $log(t)=6.6$ up to $log(t)=10.0$, with a step as $\Delta$$log(t)=0.01$, respectively. The isochrone fitting process iterates through various models of age, metallicity and extinction to find the best fit, ensuring accurate estimates of the cluster metallicity.

We have controlled the accuracy of the metallicity found from variable stars with this code. We assumed that the metallicity of the cluster members is the same.
According to \texttt{MetalCode}, we found the best fit for the cluster fundamental parameters as follows: $d=832$ pc corresponding to distance modulus $DM_{0}=9.6$ mag, $E(G_{BP}-G_{RP})=1.10$ mag corresponding to $E(B-V)=0.85$ mag, $t =200$ Myr, and $Z=0.005$.

Metallicity values for 110 member stars were provided by GaiaDR3, with a median value of $[Fe/H]=-0.57$. This value is very close to the metallicity obtained using \texttt{Metalcode} ($[Fe/H] = -0.54$). Furthermore, we constrained the cluster metallicity through the variable stars, determined by three different methods, as seen in Table~\ref{tab:specphoto_variable_table}. However, for the reasons outlined below, our attention was directed towards the metallicity of three pulsating stars situated in close proximity to the blue part of the main sequence (MS), accurately representing the single stars within UPK~220.
As seen in Table~\ref{tab:specphoto_variable_table}, although ID 16 has been identified as an evolved star ($logg \simeq 3.95$), its brightness variations do not align with the evolved region of the isochrone. This issue, as discussed later in Section~\ref{sec:resvar}, raises some concerns about the membership of the star in the cluster. On the other hand, the metallicity values of ID 42 obtained from three methods are different from each other (see in Table \ref{tab:specphoto_variable_table}). Due to these uncertainties, IDs 16 and 42 were excluded from consideration when determining metallicity.
Since the metallicities obtained from GSP-Phot analysis are reliable and consistent with the error limits, the metallicity of three stars (IDs 49, 138 and 147) were considered in determining the input metallicity of MIST isochrones applied to the cluster.
Thus, the average metallicity derived from these three variable single stars was calculated as $[Fe/H] = -0.56$, with $Z = 0.004$.

\subsection{Determination of Physical Parameters}

\begin{figure*}
\centering
\includegraphics[width=8 cm, angle=270]{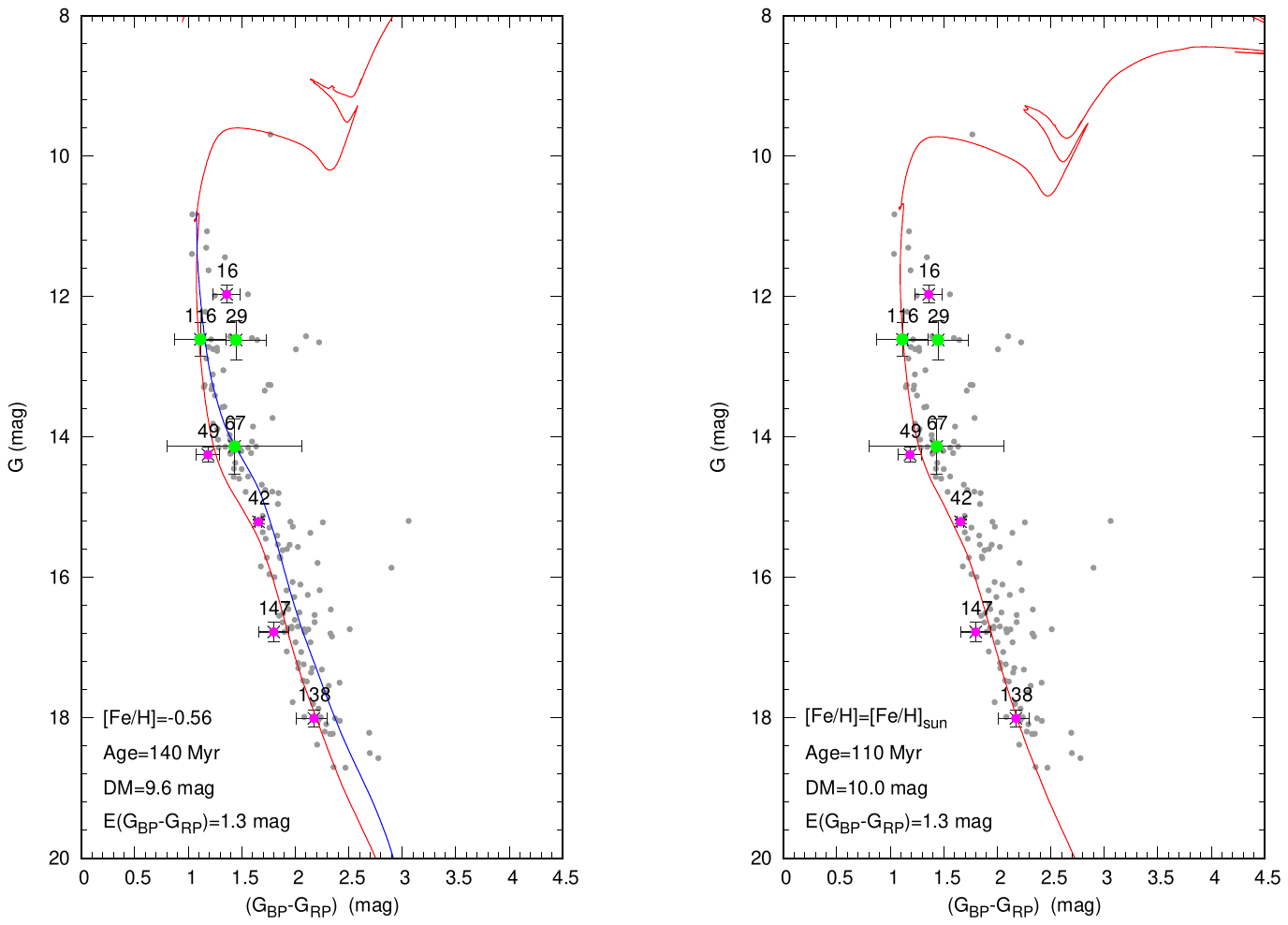}
\caption{The observational $G$ versus ($\rm G_{BP} - \rm G_{RP}$) CMDs of UPK~220 from GaiaDR3. The isochrones fitting is done via MIST considering two metallicities (left panel: $[Fe/H]=-0.56$, right panel: solar abundance). The member stars are described as grey dots. Magenta circles show eclipsing binary systems, whereas the green circles correspond to the single variables. The blue isochrone in the left panel represents the photometrical effects of the binary stars ($q=1$) on the CMD.}
\label{fig:cmd1}
\end{figure*}

Despite metallicity being one of the four free parameters typically considered when fitting isochrones to the cluster MS, it is often ignored or assumed to be solar abundance in many studies. 
Consequently, an unknown bias is introduced in the estimation of the age, reddening, and distance of the cluster \citep{2010A&A...517A..32P}. 
Therefore, to demonstrate the effect of two different metallicities on the cluster parameters in the CMD of UPK~220, the results were examined in terms of solar abundance and the metallicity found by variable stars, as shown in Figure~\ref{fig:cmd1}. Here, the red solid lines in the figure represent \textit{MIST} isochrones with $[Fe/H]=-0.56$ (left panel) 
and solar abundance (right panel). Gray points on the diagram represent potential member stars, while stars marked in magenta and green circles represent the member variable stars classified as eclipsing binaries and the others, with their IDs listed in Table~\ref{tab:variable_table}.

Our fitting procedure to determine cluster parameters from the CMD was as follows:
Initially, we  applied a CMD fit to all members assuming the solar metallicity (see in Figure~\ref{fig:cmd1}, right panel), without including the metallicity constraints as given in Section~\ref{metal}. However, by considering the metallicity obtained from the variable stars and \texttt{Metalcode}, the \textit{MIST} isochrone fit was reapplied to best represent the CMD of all member stars. After the constraining the metallicity, we obtained the age and distance of the cluster as shown in Figure~\ref{fig:cmd1} (left panel). We then compared the cluster parameters obtained from the variable stars and found them to be consistent with each other. These consistencies are discussed in the following section.

Based on the initial metallicities for the isochrones, the age and reddenings of the cluster were estimated as 140 Myr and $E(G_{BP}-G_{RP})=1.3$ mag for $[Fe/H]=-0.56$, and 110 Myr and $E(G_{BP}-G_{RP})=1.3$ mag for $[Fe/H]=[Fe/H]\odot$, respectively. These reddenings correspond to $E(B-V)=1.0$ mag for both metallicities, determined using the reddening ratio from \citet{2018A&A...619A.176B}. 
The isochrones were constructed with the intention of passing through the same turn-off point at both metallicities, and were fitted to provide the optimal fit. 

While the differences in metallicity do not cause a significant difference in age and reddening, they do affect the distance modulus, with ${DM}_{0}=9.6$ mag for $[Fe/H]=-0.56$ and $DM_{0}=10.0$ for $[Fe/H]=[Fe/H]_{\odot}$. 
This difference leads to a distance discrepancy of nearly 200 pc 
and affects parameters such as mass (M), $logg$, 
and $T_{\rm eff}$ obtained from 
the isochrone fit for the member variable stars. 
The parameters for the variable stars obtained 
for $[Fe/H]=-0.56$ can be seen in Table~\ref{tab:specphoto_variable_table}.

Since the effect of stellar rotation on the MS and, consequently, on the age is 
not observed in young clusters (40 to 300 Myr) \citep{2013ApJ...776..112Y, 2021JApA...42...60A}, this effect has been neglected 
for one of the young cluster UPK~220, 
and rotation has been disregarded in the isochrones.

\section{Analysis Results of Member Variable Stars}
\label{sec:resvar}

The brightness changes of member variable stars obtained from the TESS analysis 
are represented in Figure~\ref{fig:cmd1} as error bars in both the G and $(G_{BP}-G_{RP})$ planes. The binary sequence (BS) 
for equal mass systems ($q=1$) is 
shown as a blue solid line in the left panel. 
Accordingly, except for a few 
stars (IDs 16 and 29), the majority of the member variable stars 
approach the MS within these 
brightness limits or fall between the MS and the BS. 
The stars with IDs 16 and 29, which could 
potentially be cluster members, 
do not coincide with the cluster MS and BS. Although ID 16 shows a deviation from 
the cluster MS, the position of 
ID 29 coincides with the MS and BS within the brightness change limits. 
This raises questions about the accuracy of the age determined for the cluster 
or whether this star 
is a member of the cluster. 
If these stars (IDs 16 and 29) are indeed potential 
members of the cluster, the cluster age should shift towards older ages. 
However, the presence of 
the stars that are well fitted with the 140 Myr isochrone above 
the member variable stars and one 
evolved star located in the subdwarf region points to younger ages.

\subsection{Binaries}

The dynamical evolution of a star cluster is heavily influenced by its binary population. Even a 
small initial fraction of binary stars can have a considerable impact on cluster dynamics 
and the overall 
evolution of the stellar population of the cluster. 
Since binary interactions (soft and hard) regulate internal energy and drive mass segregation, their existence influences the dynamical stability and lifetimes of OCs. The number of binary fractions, types of eclipsing binaries (detached, semi-detached and contact), and evolution of binary stars may have significant impacts on the dynamical evolution of OCs \citep{goodwin2010binaries, de2015dynamical, piatti2017highly, wang2022impact}.

Although an open star cluster contains a large number of binary systems, detecting all of them is quite challenging. According to \citet{CHEN2024104083}, approximately 50\% of the member stars in clusters are expected to be binaries. It is evident that photometric data of Gaia is highly precise, \citet{2025arXiv250101617L} found that binaries with 
$q < 0.5$ are difficult to detect using optical data alone. Similarly, despite \citet{2021ApJ...921..117T} having spectroscopic data spanning 39 years, the detection of lower-mass binaries remains constrained due to uncertainties in orbital periods and orbital inclinations. 

In addition, due to the availability of photometric TESS data, only eclipsing binaries with short periods (up to $\sim$27 days) can be discovered. Furthermore, due to the infrared sensitivity of TESS data, it is highly likely that many of these binaries remain undetected. In this study, considering the number of member stars in UPK~220, at least 70 stars are expected to be binaries. However, due to the fact that mentioned above regarding TESS, only three binary stars have been detected.

It is commonly understood that an unresolved binary system, consisting of two identical stars, 
exhibits the same color but twice the luminosity compared to a single star with equivalent 
properties. Additionally, such a system, composed of two MS stars with $q=1$, will 
manifest a vertical displacement of 0.753 mag in the cluster CMD, 
regardless of the wavelength bands utilized \citep{1998MNRAS.300..977H}. A system with two unequal 
MS components will exhibit a combined color that is redder than that of the brighter 
component. Its luminosity is higher than that of a single star but lower than that total luminosity of an equal-mass 
binary \citep{1998MNRAS.300..977H}.  
The difference in colors and magnitudes is influenced by the mass ratio 
$q=M_1/M_2\leq1$ where $M_1$ and $M_2$ represent the masses of the primary and secondary 
components, respectively. As 
$q$ approaches zero, the position of the binary will converge towards the MS, since the impact of the 
fainter component becomes negligible \citep{2024AJ....167..100Y}. 

As mentioned above, the positions of the binary stars in the CMD (as shown with green circles in Figure~\ref{fig:cmd1}) 
may vary depending on the mass ratios of the system and their respective contributions to the total 
radiation. Since the parameters are determined based on the total 
brightness of the binary, the results may be misleading depending on the contributions from each 
component. Additionally, the \textit{Gaia} brightness given in the CMD may vary depending on the phase of the 
binary system at the time of observation. Since the \textit{Gaia} database provides average brightnesses for 
any star, it is inevitable that brightness differences will be observed for the 
variable stars between 
observations. Therefore, instead of \textit{Gaia} magnitudes, the $T_A$ and $T_B$ magnitudes 
of each component obtained from \textit{TESS} light curves were used in this study. 
Converting TESS to \textit{Gaia} bandpasses, we use the relation in 
\citet{2019AJ....158..138S}. 

\begin{figure*}
    \centering
    \begin{minipage}{0.6\columnwidth}
    \includegraphics[width=\columnwidth]{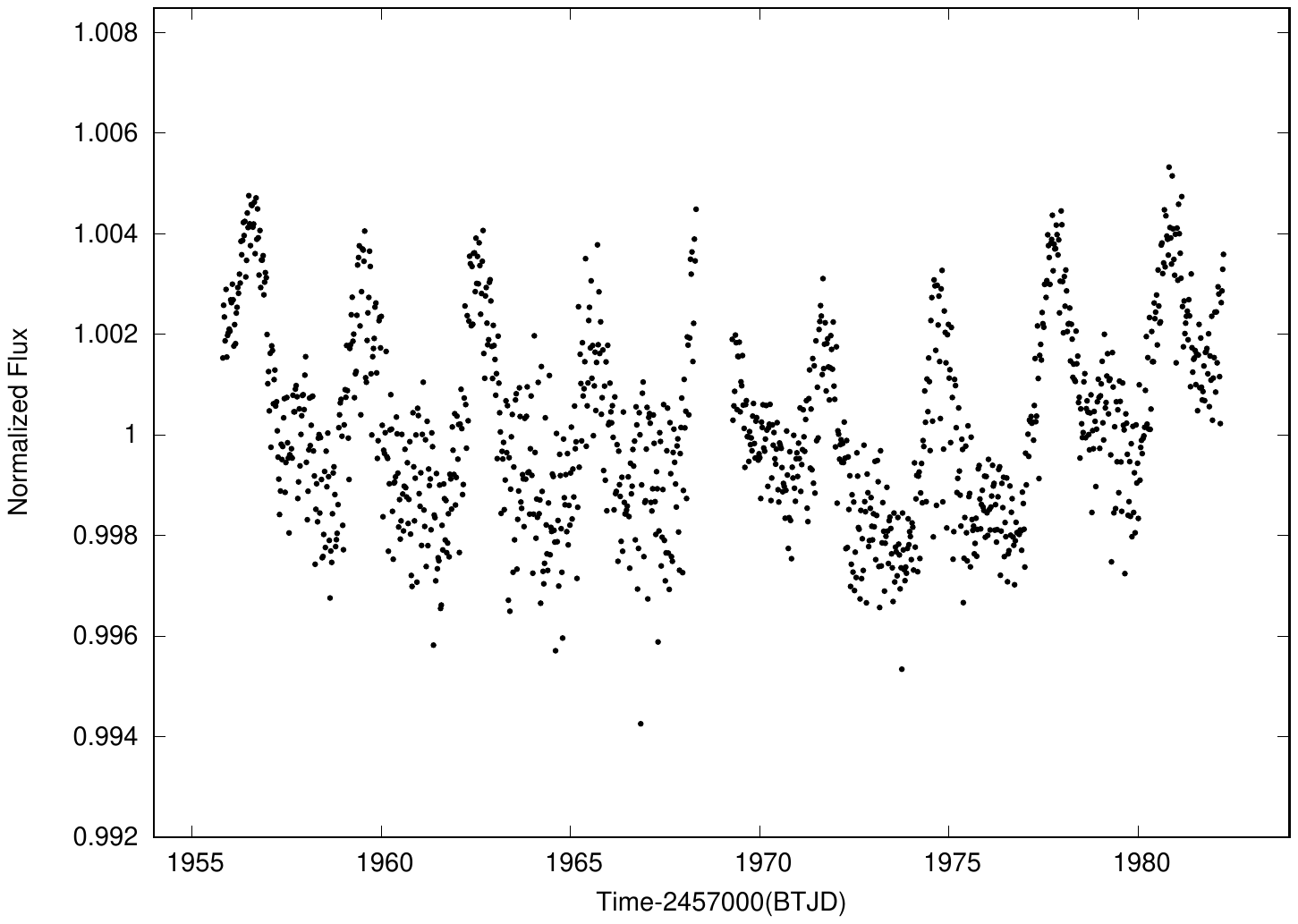}
    \end{minipage}
    \begin{minipage}{0.6\columnwidth}
    \includegraphics[width=\columnwidth]{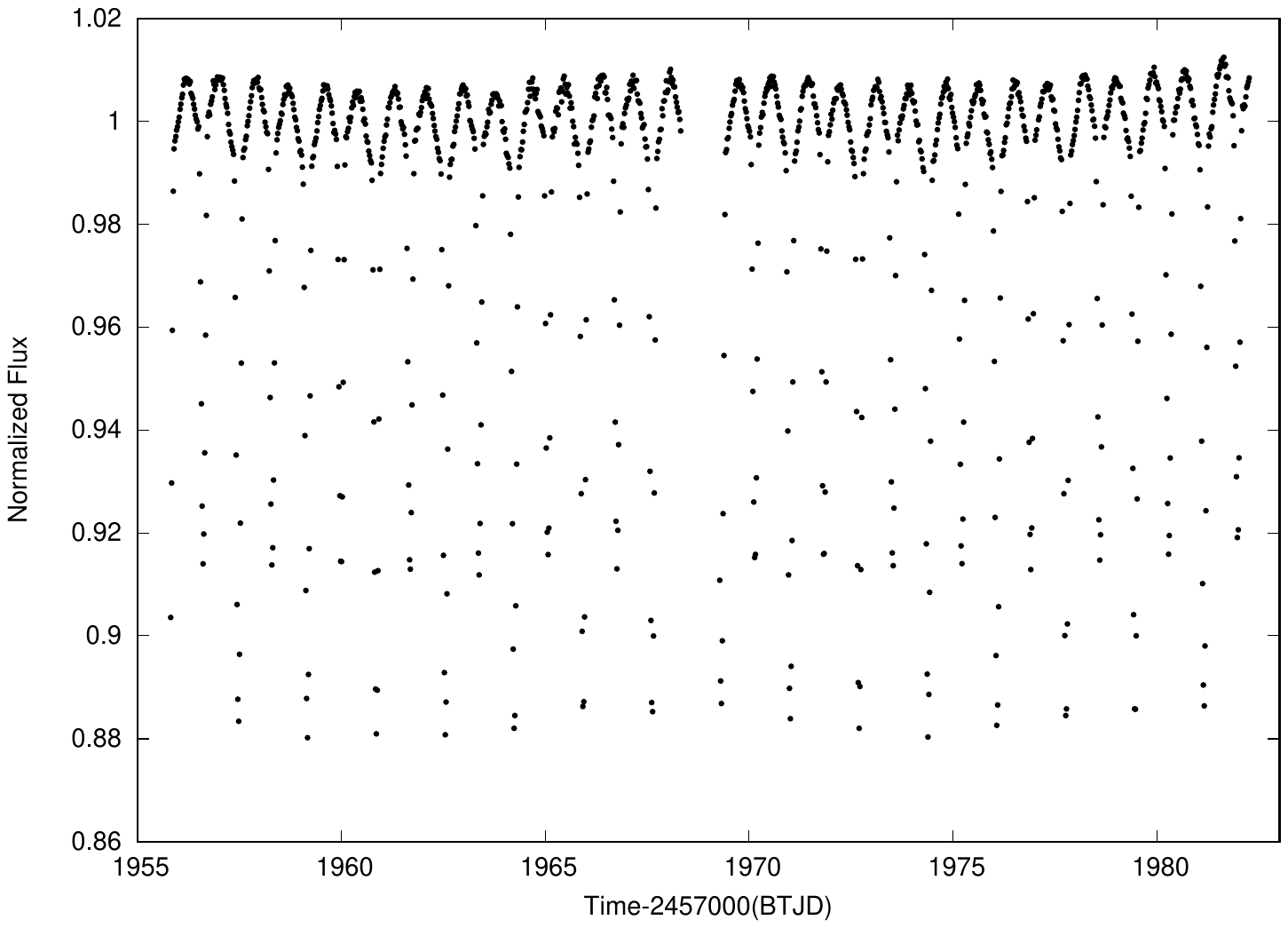}
    \end{minipage}
    \begin{minipage}{0.6\columnwidth}
    \includegraphics[width=\columnwidth]{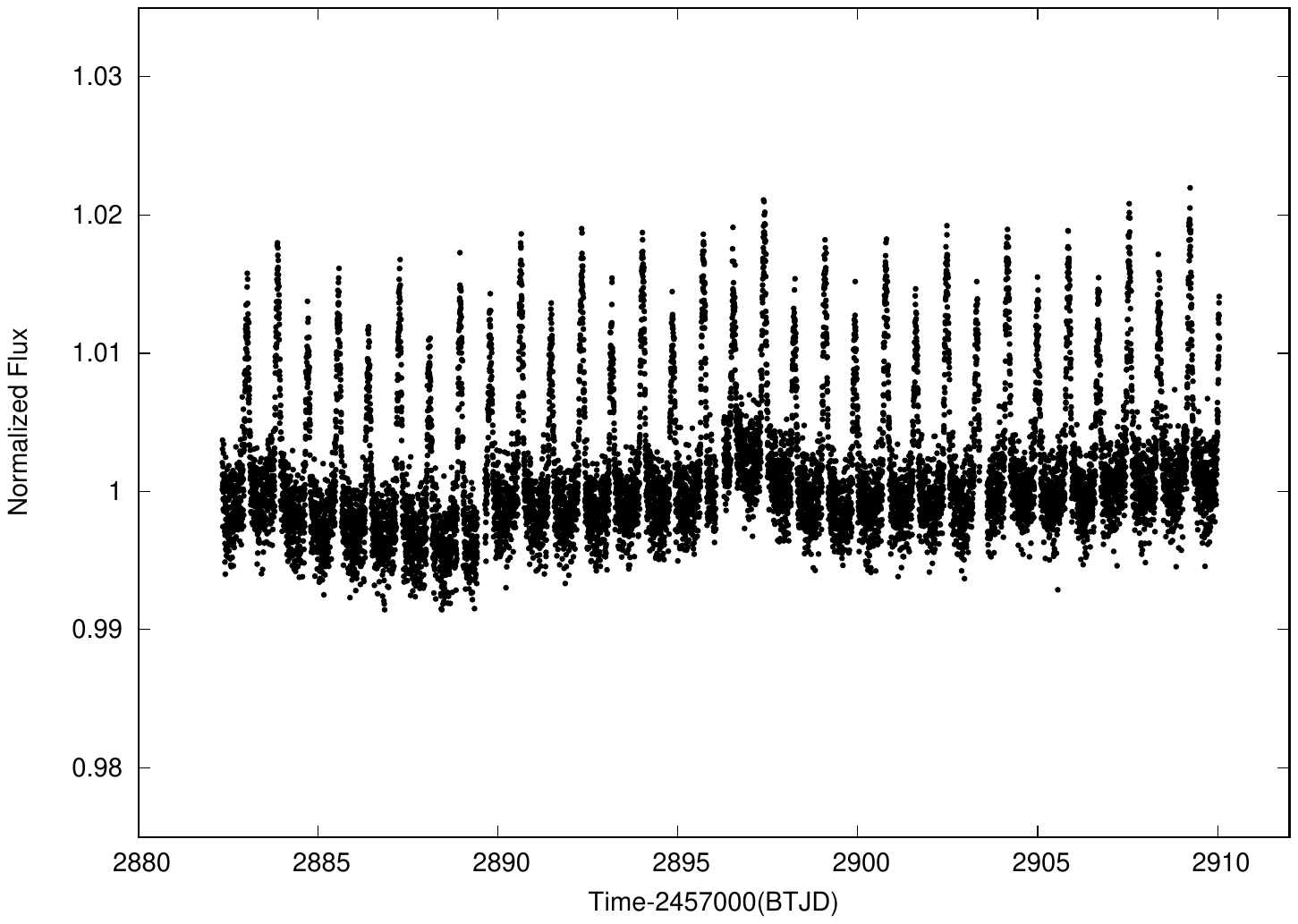}
    \end{minipage}
    \begin{minipage}{0.6\columnwidth}
    \includegraphics[width=\columnwidth]{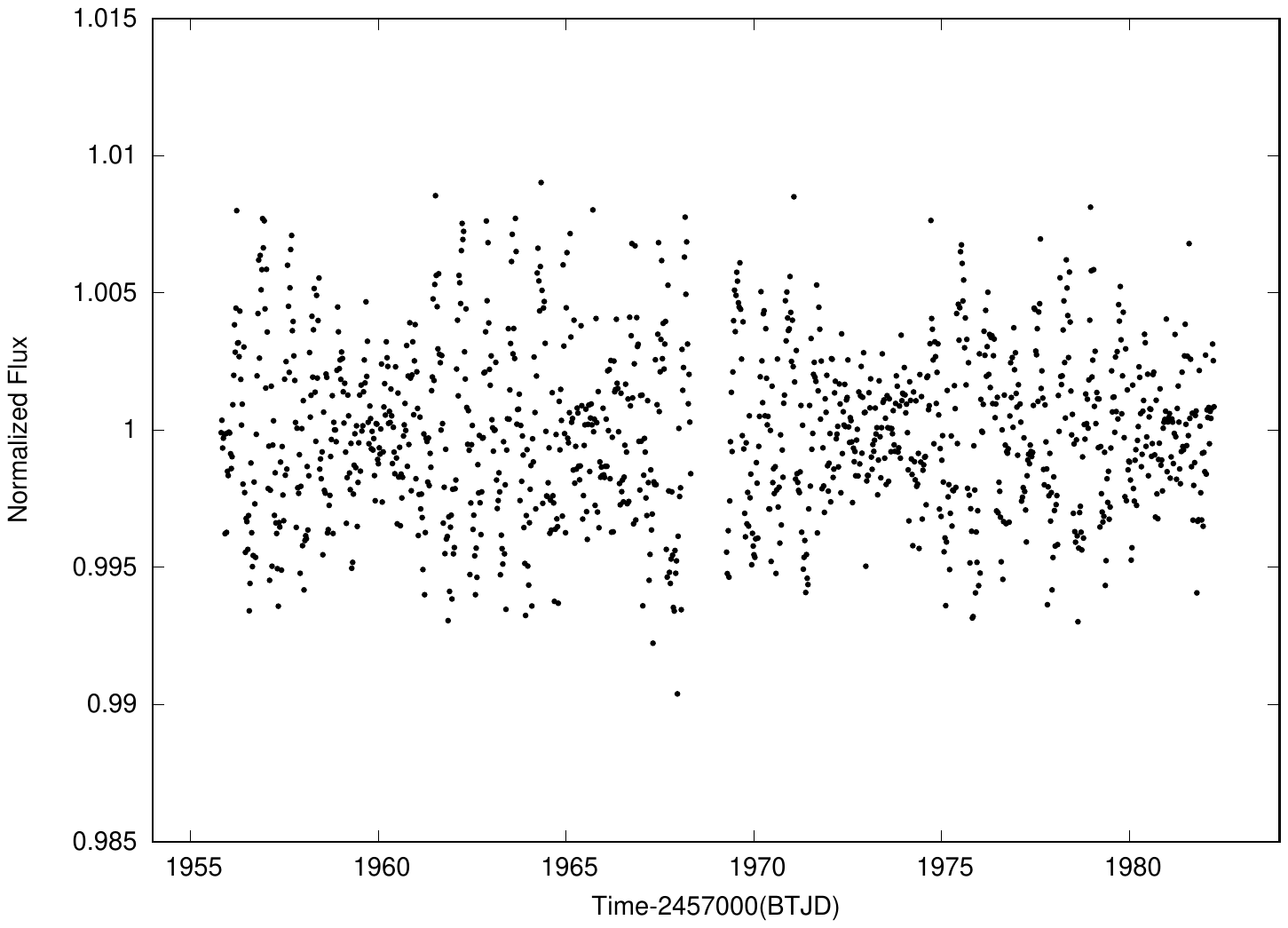}
    \end{minipage}
    \begin{minipage}{0.6\columnwidth}
    \includegraphics[width=\columnwidth]{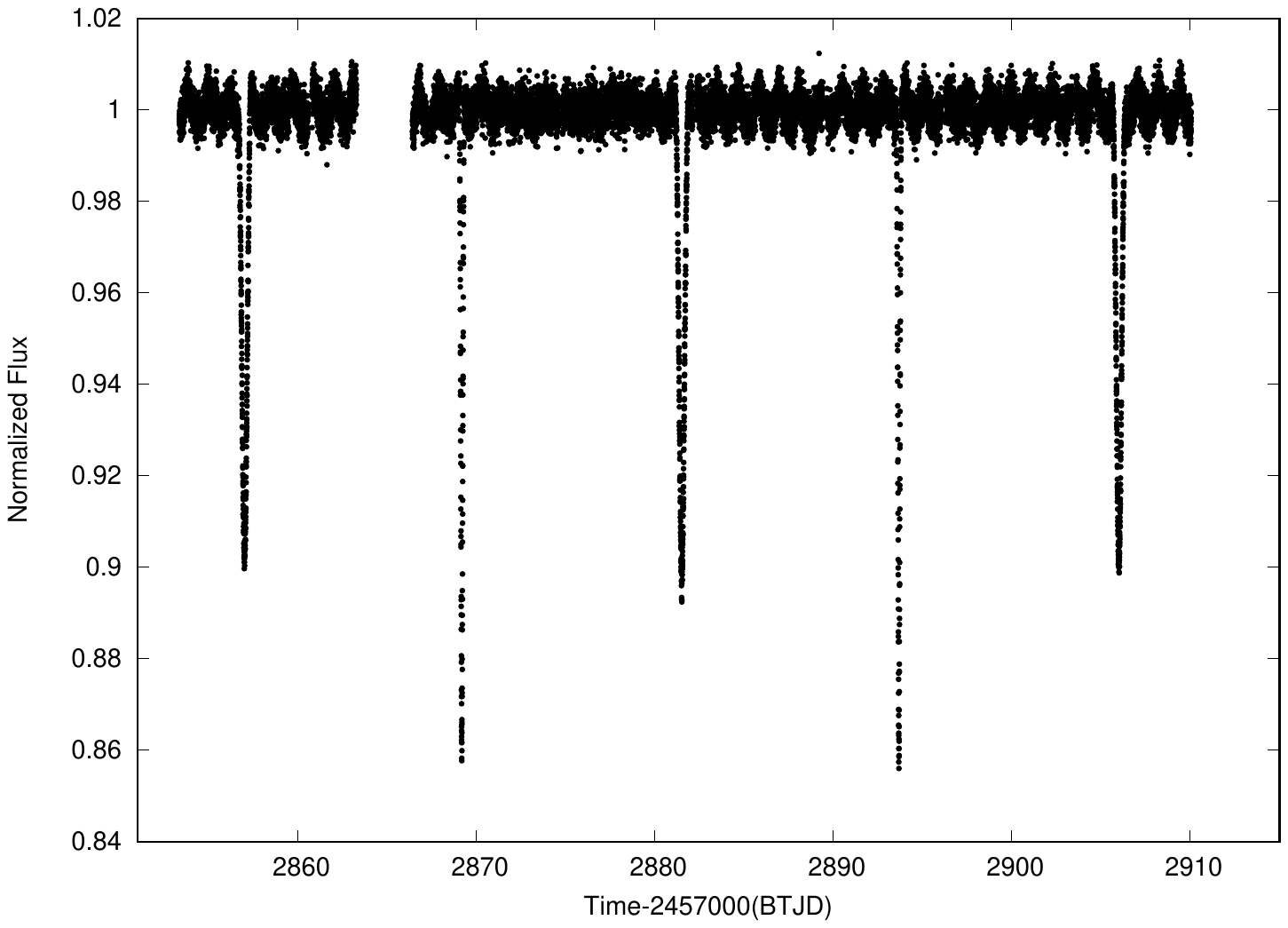}
    \end{minipage}
    \begin{minipage}{0.6\columnwidth}
    \includegraphics[width=\columnwidth]{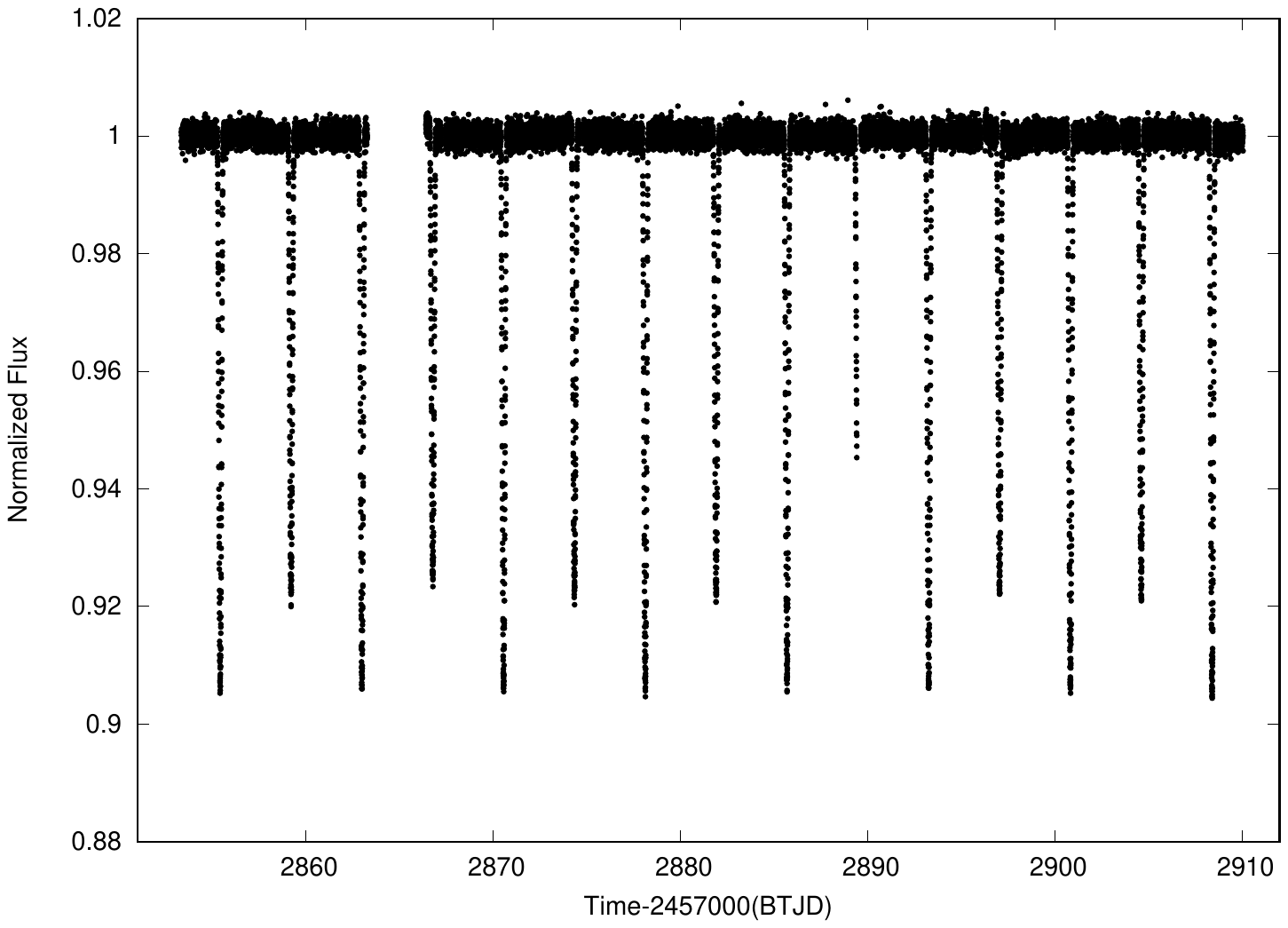}
    \end{minipage}
    \begin{minipage}{0.6\columnwidth}
    \includegraphics[width=\columnwidth]{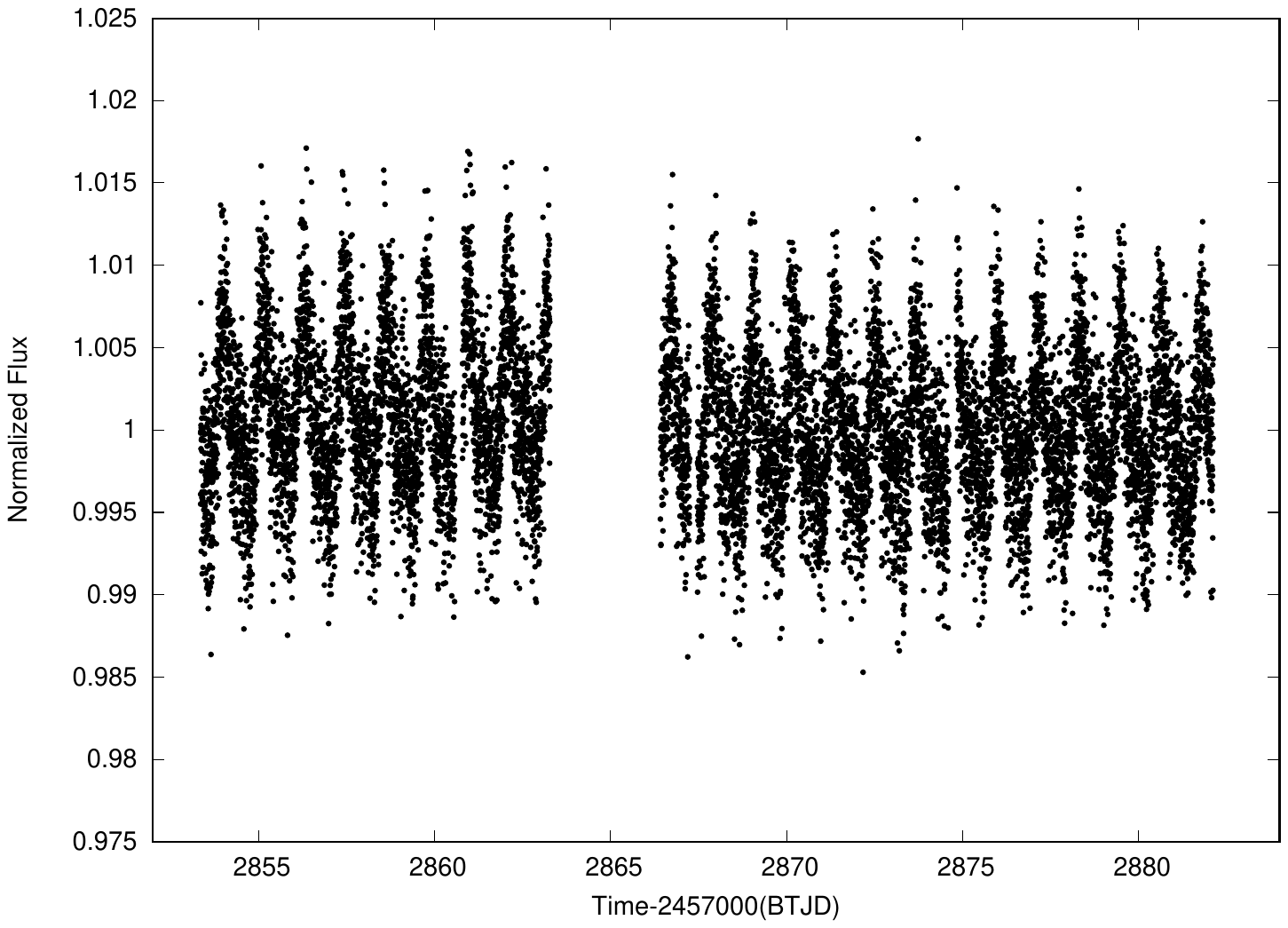}
    \end{minipage}
    \begin{minipage}{0.6\columnwidth}
    \includegraphics[width=\columnwidth]{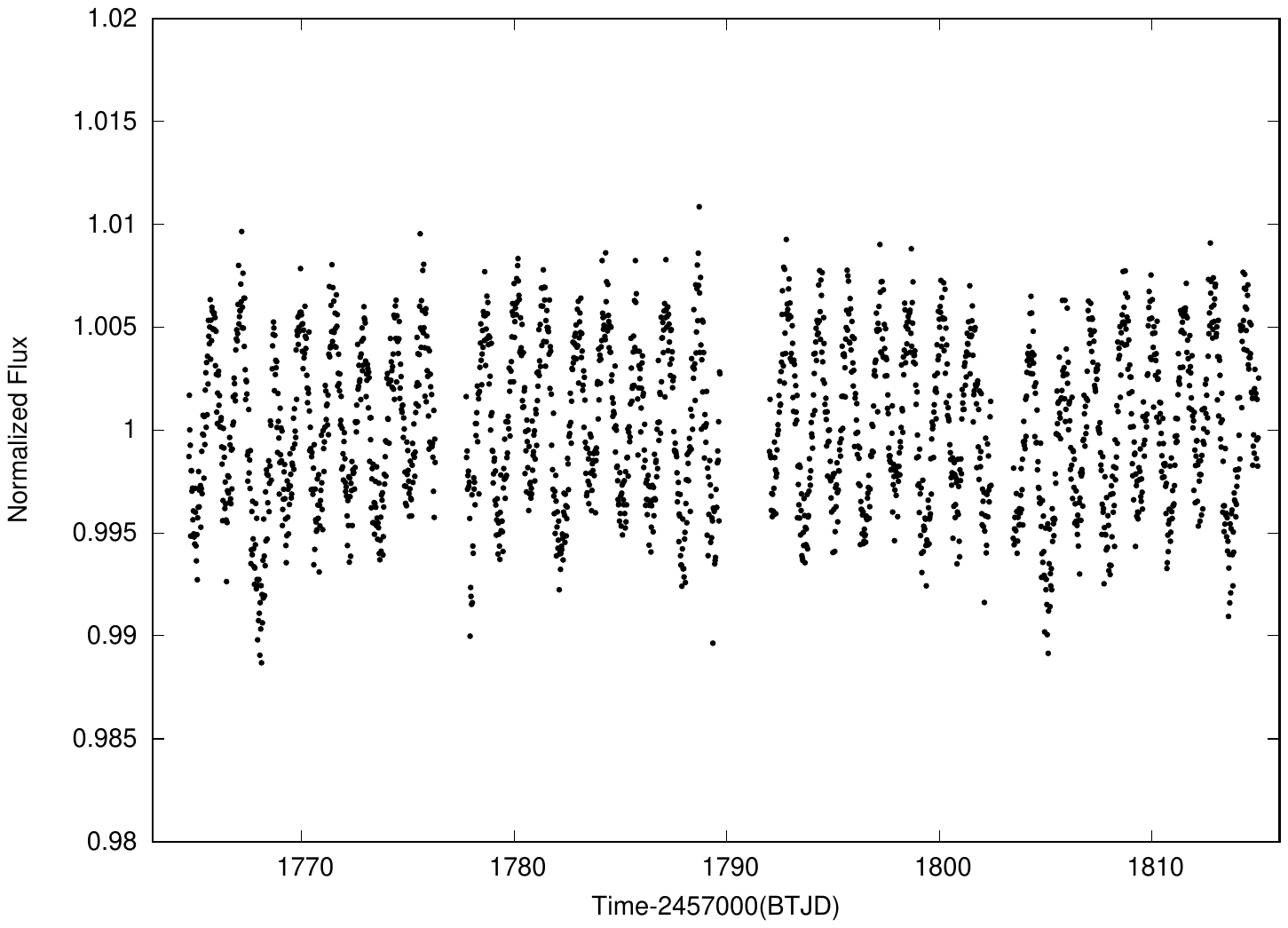}
    \end{minipage}
    \caption{Normalized TESS light curves of variable member stars. First row: ID16 (Sector 24), ID29 (Sector 24) and ID42 (Sector 58). Second row: ID49 (Sector 24),
    ID67 (Sectors 57 and 58) and ID116 (Sectors 57 and 58). Third row: ID138 (Sector 57) and ID147 (Sectors 17 and 18).}
\label{fig:lcs}
\end{figure*}

When we check the positions of these stars on the CMD, 
IDs 29 and 67 have 
abnormal positions and are displaced from the MS. 
Considering the phases of these stars and their 
positions to the MS, comparisons of these stars with the model results are as follows:

\begin{itemize}

    \item \textbf{ID 29}: The system is a detached binary and in addition to 
primary and secondary eclipses, 
ellipsoidal effect is also clearly seen as a sinusoidal 
variation in the light curve of the star (see in Figure~\ref{fig:lcs}).
In the light curve of the star, 
six primary minima (\textit{from 2458764.5 to 2458765.0 BJD in sector 17,
from 2458802.0 to 2458803.5 BJD in sector 18, 
from 2458955.0 to 2458956.0 BJD and from 2458968.0 
to 2458969.5 BJD in sector 24, from 2459853.0 to 24589853.5 BJD in sector 57}) 
and 
three secondary (\textit{from 2458968.0 to 2458969.5 BJD in sector 24, 
from 2459903.4 to 2459903.6 BJD
and from 2459910.0 to 2459910.1 BJD in sector 58}) appear distorted. 

On the other hand, there is no data for the out-of-eclipse portion of the 
\textit{TESS} light curve of ID 29 during the following intervals: 
from 2458791.0 to 2458791.5 BJD in sector 17, from 2459860.5 to 
2459861.0 BJD in sector 57 and 
from 2459882.0 to 2459882.5 BJD, from 2459889.0 to 2459890.0 BJD and
from 2459896.0 to 2459896.5 BJD in sector 58. 

For the light curve models of ID 29, 
we assume circular orbit 
because the system has orbital periods 
lower than 5 days \citep{2013AJ....145....8G}. 
According to the mass values of $M_A$ and $M_B$ obtained 
from \textit{MESA} binary models, the mass ratio was 
calculated to be $q=0.6$. 

When the GSP-Phot and SED values based on the total brightness are compared to the MESA binary model, 
there are differences in temperature. While GSP-Phot and SED temperatures are consistent with each other 
($\sim$ 9500 K), the MESA model yields $T_{\rm eff,A}=11500$ K and 
$T_{\rm eff,B}=8900$ K for $M_A$ and $M_B$, respectively. 
However, when considering the contributions of 
both components to the system temperature, 
these values fall within the effective temperature 
ranges derived by GSP-Phot and SED, 
indicating that reasonable results are obtained from MESA binary model. 

\item \textbf{ID 67}: 
The light curve of ID 67 represents a typical detached eclipsing binary
system containing a pulsating component.
In TESS observation of the star, data gaps of totally $\sim$1.5, 
$\sim$1.4, $\sim$1, $\sim$4 and $\sim$1 days 
were encountered in sectors 17, 18, 24, 57 and 58,
respectively. Nevertheless, it has been noticed that 
none of these data gaps impact
primary and secondary minima in our analysis. 

A total of five primary minima
and six secondary minima were analyzed in this study.
We have clearly seen the variations 
in the residuals of the fits (see in Figure~\ref{fig:lcs}). 
We therefore perform a Lomb-Scargle analysis 
for the star and discover Cepheid-like pulsations with 
a period of $\sim$2.5 days. 
The derived pulsating frequencies with their amplitudes, phases and 
estimated periods of each sector TESS observations are 
presented with their uncertainties in  
Table~\ref{tab:tableper3}.  

We consider the light curve brightness change behaviour 
and period of the system
that the companion is a classical Cepheid($\delta$ Cephei-type). 
Classical Cepheids are intermediate-mass, young stars \citep{2021ApJS..253...11P}.
In this respect, our MESA model results for 
ID 67 are consistent in terms of age and mass.
However, spectral observations are essential to achieve more precise results.

The presence of a Cepheid component in ID 67 allowed for a comparison of the GaiaDR3 distance. 
We calculated the distance of the star 
as 987 $\pm$ 37 pc using Gaia parallax and 1148 $\pm$ 205 pc using 
the period-luminosity relation in \citet{2022ApJ...927....8O}. 
This discrepancy in distance may arise from factors such as metallicity, parallax offset, orbital parallax, or the effects of binary evolution on Cepheid variables. For example, when orbital parallax is included, according to the calculations of \citet{2023A&A...669A...4G}, there is approximately a 5\% change in the GaiaDR3 parallax. Under this assumption, the distance of ID 67 is calculated to be between 924 and 1060 pc.

As shown in Table \ref{tab:magbin}, $T_{eff}$ and $logg$ values obtained from GSP-Phot and SED analyses for the components A and B are consistent, indicating a similar evolutionary stage. However, the MESA model results reveal that the higher-mass component is hotter than the lower-mass component. For the metallicity, 
the most of the effective temperature contribution in the SED and GSP-Phot analysis comes from the high-mass component.

Examining the CMD position of the system, the Gaia brightness places it in the BS. However, due to the large brightness variations, estimating 
mass ratios directly from the BS is challenging. Based on the TESS brightness variations and the mass values derived from models, 
there is an approximate twofold difference in mass between the components, explaining the significant brightness variation.

Additionally, the presence of pulsation in the system causes the 
brightness variation to be very large, as seen in Figure \ref{fig:lcs}. Furthermore, the very similar 
results found in the two MESA models reduce the margin of error in mass values, and this is supported 
by other parameters. 
Considering the mass ratio of $q=0.5$, this indicates that the B component is 
more evolved. 
Indeed, the $logg$ value supports that the B component is more evolved.

\item \textbf{ID 116 }: 
The system is a detached eclipsing binary. 
The TESS light curve of the star, derived from sectors 17, 18, 24, 57 and 58, contains a
total of 18 primary and 14 secondary minima but unfortunately, 
three secondary minima have incomplete or data gaps.  
Additionally, the out-of-eclipse data of ID 116 observed in five different
sectors in total exhibits an intermittent data gap of $\sim$11 days.

We conducted Lomb-Scargle analysis of the residuals for the star and 
found low-amplitude pulsation with a period of $\sim$5.5 days,  
observed only in the TESS data for sectors 57 and 58.
The derived pulsation frequencies, 
along with their amplitudes, phases and estimated periods from the
residuals are presented with their uncertainties 
in Table \ref{tab:tableper3}. We assumed \textit{BTSettle} 
fit for primary star and a blackbody fit for 
component stars of ID 116. 
Through SED analysis, 
the two components of the system were photometrically distinguished, and the  
$T_{eff}$ and $logg$ values were determined, separately. 
However, since $T_{eff}$ obtained from GSP-Phot is 
derived from the total brightness of the binary system, 
it is not compatible with those obtained from 
SED. Although the $T_{\rm eff}$ values 
obtained from the MESA model are compatible with each component, separately. 
There is an incompatibility between the $logg$ values obtain from SED analysis and MESA models. 
According to the mass 
values obtained from the MESA models, $q \simeq 0.67$, 
and the brightness obtained 
from GaiaDR3 is also within this range.

\end{itemize}

\subsection{Other Variables}
\label{sec:othervar}

\begin{itemize}
    \item \textbf{ID 16}: Using TESS data, we estimated
the rotation period of the star as $P_{\rm rot}=3.061 \pm 0.025$ days. 
Light curve of ID 16 exhibits
spot modulations, similar to observed in magnetic active stars. 
(e.g. HD 176330 - \citet{2018AA...616A..77B}, TIC 219234021 - \citet{2019MNRAS.487.4695S}). 
Additionally, we find another period equal to $P_{\rm rot}/2$. 
We doubt that the spots on the star are 
separated by 180$^{\circ}$ in longitude (aka 
AU Mic-like effect - \citet{2024MNRAS.528..963M}). 

Using gyrochronology relation from \citet{2007ApJ...669.1167B}, 
we determined the stellar age to be 
136 $\pm$ 21 Myr using $P_{\rm rot}$ of the star.
The star is located at 
a point very far from the MS in the CMD and does not approach the MS, even when 
the brightness change is taken into account. 
Despite this, all parameters found in the SED and GSP-Spec 
analyses were compatible with each other. 

\item \textbf{ID 42}: The light curve of ID 42 was derived using 
TESS observations from sectors 17, 18, 24, 57 and 58.
Similar to ID 16, the star displays spot modulations that are evident 
in the TESS light curve. 
We estimate the rotation period of the star as 
1.691 $\pm$ 0.008 days
by using Lomb-Scargle period analysis. 

However, the gyrochronology relation of 
\citet{2007ApJ...669.1167B} is
unable to accurately determine 
the stellar ages with $P_{\rm rot} \leq 2$ days. 
Therefore, we estimate the minimum age of the star to be 50 Myr.
In the CMD, the star is located close to the MS, 
and its brightness variation as a variable star is relatively small. 
The mass value obtained from the MESA model is approximately $M=1.40 {M_\odot}$. 
The $T_{\rm eff}$s derived from 
the GSP-Phot ($\sim8000$ K) and SED ($\sim7500$ K) 
analyses are very close to each other, although they 
are slightly lower than obtained from the MESA model ($\sim8500$ K).

\item \textbf{ID 49}: The star exhibits a classical $\gamma$ Dor 
behavior in the TESS light curve.
We identified gravity (g-) mode pulsation of ID 49 through 
Lomb-Scargle periodogram analysis.
The fundamental period was determined to be 0.6663 $\pm$ 0.0028 days.
The pulsation frequencies of the star are listed in Table~\ref{tab:gammador}. 
Although g-mode pulsations can be used to estimate the age of $\gamma$ Dor stars \citep{10.1093/mnras/stz501}, the TESS data of ID 49 are insufficient to obtain g-mode period spacings. This issue can be addressed in the future with long-term continuous observations.

$\gamma$ Dor stars, one of the well-known 
non-radial pulsating variables, fall within the spectral-type range 
between late-A and early F-type stars. They exhibit a characteristic light 
curve variations and 
their period range varies between $\sim$0.3
and $\sim$3.5 days \citep{1999MNRAS.309L..19H}. 
The light curve variation and fundamental pulsation period range indicates that
ID 49 is a $\gamma$ Dor variable candidate. 
Considering the classical instability strip, $\gamma$ 
Dor stars are typically found within the range of 7000-10000 K \citep{2005AA...435..927D}.
The $T_{\rm eff}$ of ID 49, determined as $8577^{+74}_{-82}$ 
from GSP-Phot and 
$8353^{+71}_{-89}$ from SED analysis, 
supports its classification as a $\gamma$ Dor variable.

Previous studies have not identified 
$\gamma$ Dor variables in OCs
older than $\sim$250 Myr \citep{1999IBVS.4705....1K}. 
Therefore, it provides further evidence
that the age of the UPK~220 cluster is less than $\sim$250 Myr, 
consistent with findings in this study. 
Nevertheless, there are also studies in the literature that present evidence
to the contrary. Among these, a comprehensive study that also takes $[Fe/H]$ of
clusters into consideration was conducted by \citet{2009AcA....59..193M}.
They found that the relation between cluster 
age and $[Fe/H]$ of an OC and 
the probability of observing a $\gamma$ Dor variable in an OC
is not dependent on the cluster age. Furthermore, they result that
the probability of observing a $\gamma$ Dor increases with $[Fe/H]$ 
in the cluster after a certain age as $log(age)\simeq8.5$. 
However, they also emphasise that the membership
of $\gamma$ Dor stars discovered in these OCs should be carefully
reconsidered. 

Upon examination of the studies about $\gamma$ Dor stars in OCs
\citep{1995AJ....110.2813P, 2000ASPC..198..225G, 2003AJ....125.2085S, 2003ASPC..294..379B, 2004A&A...422..951C, 2008AA...489..403P, 2009A&A...493..309S, 2012MNRAS.419.2379J},
it was determined that UPK~220 is one of the two clusters (the other is NGC 581 -
\citet{2003NewA....8..737T}) 
with a metal-poor $\gamma$ Dor member among the Galactic OCs exhibiting such
variables.

As shown in Table~\ref{tab:magbin}, 
all $T_{\rm eff}$ values 
obtained from the GSP-Phot and SED analysis, and MESA models 
are compatible with each other.
The mass value was found to be $M=1.40$ ${M_\odot}$. This mass is 
consistent with the range of $\gamma$ Dor mass 
($1.4-2.0$ $M_{\rm \odot}$ - \citet{2020AA...635A.106T}).

\item \textbf{ID 138}: The light curve exhibits a periodic variation with a period 
of 1.15926 $\pm$ 0.00166 days
and the amplitude of the variation is roughly 1\%. 
The position of the star on CMD, the period 
between 0.01 and 5 days, 
and the amplitude of the variation \citep{2006PASP..118.1390P}
indicate that the star is a T-Tauri
candidate. However, the lack of observational data in the infrared region of the 
spectrum precludes a definitive conclusion 
regarding the red excess of the star, as determined 
on SED.

\citet{1994AJ....108.1906H} divides T-Tauri stars into three subtypes based on 
their photometric variations. ID 138 appears to belong to the Type I class as it
exhibits cyclic variations with low amplitude modulation.
\citet{1994AJ....108.1906H} also note that the stars of Type I class 
display rotational modulation due to cool
spots and have no visible accretion disks in their spectroscopy. 
However, due to the lack of spectroscopic observations of ID 138, 
it is not possible to reach a definitive conclusion on this issue.

The star is located in the faint part of the MS in the CMD, and its mass, derived from the MESA model, is estimated to be $M=0.70$ ${M_\odot}$. 
The $T_{\rm eff}$s from the GSP-Phot and SED analyses 
are lower than those predicted by the MESA model, with the 
$T_{\rm eff}$ range of the star varying between 4400 and 5300 K.
The relatively low mass and cooler temperature compared to the Sun 
further support the classification of ID 138 as a T-Tauri star candidate.

\item \textbf{ID 147}: Upon analysis of the TESS light curve of the star, 
similar to ID 138, it
may be a potential Type I class T-Tauri star candidate within UPK~220. 
This conclusion is based on the estimated period and 
amplitude of the variation ($\sim$ 2 mmag).  
The light curve of ID 147 exhibits spot modulations 
with a period of 2.857 $\pm$ 0.004 days.
Utilising the rotation period of the star in the gyrochronology relation 
from \citet{2007ApJ...669.1167B}, 
we estimate the stellar age to be 194 $\pm$ 56 Myr. 

Due to the lack of observational data in the infrared region,
it is not possible to confirm the presence of an accretion disk 
based on the SED of the star. It is evident that further spectral observations are 
necessary to substantiate the presence of the disk around the star.
The star is low-mass, and its position in the CMD, 
located in the fainter region of the MS,
is compatible with its mass $M=0.80$ ${M_\odot}$ 
that is derived from MESA model. The parameters
derived from GSP-Phot and SED analysis, 
and MESA also yielded comparable results (see in 
Table~\ref{tab:specphoto_variable_table}). 

\end{itemize}

\section{Discussion}
\label{sec:diss}
 
We assume that stars with membership probabilities 
$P \geq 0.725$  are the most probable members of UPK~220.
Using the probability range, we derived mean proper motions 
as ${\mu_{\rm \alpha} \cos(\delta)}=-2.41 \pm 0.11$ mas/yr 
and ${\mu_{\rm \delta}}=-2.64 \pm 0.12$ mas/yr. This 
proper motions are consistent with those reported by \citet{2024A&A...686A..42H} as 
($\mu_{\rm \alpha}=-2.41$ and $\mu_{\rm \delta}=-2.47$ mas/yr). 

The structural parameters of UPK~220 
derived from the King profile fitting are as follows: 
the central stellar 
density $\sigma_{\rm_{0k}}= 0.835 \pm 0.121$ stars/{arcmin}$^{2}$, 
the core radius R$_{c}= 5.751 \pm 1.079$ arcmin, 
the residual stellar background density $\sigma_{\rm bg} = 0.674 \pm 0.023$ 
stars/arcmin$^{2}$, and the limit radius $R_{\rm lim}= 24.0$ arcmin. 
The radius is compatible with the findings of \citet{2024A&A...686A..42H} 
as $r_{\rm 50} = 0.213 \degr$ 
(corresponds to a physical size of $7.02$ pc). 
They define the cluster radius as 50 $\%$ of 
member stars within the tidal radius. Besides,
R$_{c}=6.15$ arcmin and $R_{\rm lim}=23.57$ arcmin 
that are given in \citet{2019JKAS...52..145S} are quite 
consistent with our results. 

We conclude that the number of member stars ($N_{\rm_{m}}$) 
within the limiting radius of UPK~220 is 148. This value shows 
considerable discrepancies compared to studies
in the literature: 
\citet{2019JKAS...52..145S} ($N_{\rm_{m}} = 102$), 
\citet{castro2020vizier} ($N_{\rm_{m}} = 91$), 
\citet{tarricq2022structural} ($N_{\rm_{m}} = 664$, $0.1 \leq  P \leq 1.0$),  
\citet{hunt2024improving} ($N_{\rm_{m}}= 323$, $0.45 \leq  P \leq 1.0$), and 
\citet{2024AA...689A..18A} ($N_{\rm_{m}}= 286$). These discrepancies in $N_{\rm_{m}}$ arise due to differences in the adopted probability range and the 
lower limit of photometric selection criteria within the cluster field.

Our results align with the membership estimation by \citet{2020AA.640A..1C}, except that variable star ID 138, identified in our study, is not listed as a member in their catalog.
It is important to point out 
that in Table~\ref{tab:literature}, 
we compare the results of \citet{cavallo2023parameter} that is
based only on their 84th percentile predictions. 

In Table~\ref{tab:literature}, the astrophysical parameters of UPK 220 are compared with those reported in the literature.  Since the studies \citep{2019JKAS...52..145S, 2020AA.640A..1C, tarricq2022structural, cavallo2023parameter} do not directly report [Fe/H] metallicities, we converted the provided metal abundance Z to [Fe/H] using an analytical expression from Bovy\footnote{ https://github.com/jobovy/isodist/blob/master/isodist/ Isochrone.py}. As shown in Table~\ref{tab:literature}, our estimated $DM$ and $d$, which are crucial for determining the fundamental parameters of the cluster, align well with the values reported in the literature. However, notable discrepancies exist in our age, metallicity, and reddening with respect to the literature. The aforementioned discrepancies between our results and the literature can be attributed to four factors: the type of photometric data used (GaiaDR2, GaiaEDR3, or GaiaDR3), the lower limit of the adopted photometric range, the membership selection criteria (probability thresholds), and the metallicity considered.
We infer that these discrepancies primarily result from differences between our adopted metallicity, as well as the types of isochrones employed in the analyses. To ensure the consistency of physical parameter space between the cluster and the member variable stars within the cluster region, we utilized MIST isochrones, derived from MESA models. Due to the fact that the MESA models are also considered for the member variable stars. In contrast, all age estimations reported in the literature are obtained from \texttt{PARSEC} isochrones \citep{bressan2012parsec} with different metallicities, considering the solar metal abundance.
It should be noted that this study presents the first direct determination of [Fe/H] for UPK~220, combining results from \texttt{Metalcode}, MESA models, SED analyses of the member variable star and GSP-Phot and GSP-Spec. 

\section{Conclusions}
\label{sec:conc}

In this study, the fundamental parameters of UPK~220 were determined by analysing the variable stars within the cluster, while simultaneously constraining the parameters of the variable stars using these cluster parameters. To achieve this, we combined GaiaDR3 and TESS photometric observations. Using GaiaDR3, we derive fundamental parameters of UPK~220 through membership analyses, and with TESS, we discover the member variable stars belonging to the cluster.

Although we have presented a different approach to cluster analysis using the most recent astrometric and photometric data in our study. 
Our research demonstrates the need for follow-up ground-based spectral and photometric
observations to determine fundamental parameters more accurately
for the variable member stars of UPK~220. 

\begin{acknowledgments}

\textit{Acknowledgements:} This study was supported by Scientific and Technological Research Council of Turkey (TUBITAK) under 
the Grant Number 122F364. The authors thank to TUBITAK for their supports.

\end{acknowledgments}

\begin{table*}
\centering
\caption{Properties of discovered member variable stars. The table contains 
 member IDs, GaiaID, coordinates, proper motions with their errors, parallax, 
 and member probability of these stars.}
\label{tab:variable_table}
\begin{tabular}{rccccccl} 
\hline
ID & GaiaID & RA & Dec & $\mu_{\rm \alpha}$ & $\mu_{\rm \delta}$ & $\varpi$ & Prob. \\
   & & (deg) & (deg) & (mas/yr) & (mas/yr) & (mas) & \\
\hline
16 &2210486179667643008 & 350.94541734 & 66.57117234 & -2.381 $\pm$ 0.010 & -2.733 $\pm$ 0.011 & 1.041 $\pm$ 0.010 & 0.982 \\
29 &2210564038834839552 & 351.59436002 & 66.47981201 & -2.505 $\pm$ 0.013 & -2.679 $\pm$ 0.011 & 1.074 $\pm$ 0.010 & 0.980 \\
42 &2210470241050931456 & 351.54188195 & 66.48240523 & -2.548 $\pm$ 0.027 & -2.688 $\pm$ 0.024 & 1.034 $\pm$ 0.021 & 0.978 \\
    49 &2210466908156116992 & 350.80052580 & 66.41964996 & -2.465 $\pm$ 0.015 & -2.556 $\pm$ 0.016 & 1.002 $\pm$ 0.013 & 0.977 \\
    67 &2210490341492729216 & 350.80329554 & 66.62039368 & -2.258 $\pm$ 0.021 & -2.665 $\pm$ 0.021 & 1.013 $\pm$ 0.019 & 0.973 \\
    116 & 2210482541837015040 & 350.64660260 & 66.48970822 & -2.619 $\pm$ 0.010 & -2.712 $\pm$ 0.010 & 1.063 $\pm$ 0.009 & 0.944 \\
    138 &2210487004303084544 & 351.30367105 & 66.59587931 & -2.366 $\pm$ 0.111 & -2.380 $\pm$ 0.099 & 1.057 $\pm$ 0.095 & 0.904 \\
    147 &2210274390542367616  & 351.18580700 & 66.17499712 & -2.296 $\pm$ 0.055 & -2.623 $\pm$ 0.048 & 0.990 $\pm$ 0.047 & 0.847 \\
\hline
\end{tabular}
\end{table*}

\begin{table*}
\tiny
\caption{Best fit solutions for TESS light curves of IDs 29, 67 and 116 using {\texttt{jktebop}} code. Using Monte Carlo and residual-permutation simulations, we determined the 1-{$\sigma$} uncertainties. $ecc*cos(omega)=-0.0041462148 \pm 0.0000078612$, $ecc*sin(omega)=-0.5891148379 \pm 0.0017385892$ and orbital eccentricity ($e=0.5891294283$), periastron longitude omega ($w=269.5967565681 \deg$) for ID67.}
	\label{tab:tablelcs}
	\begin{tabular}{llccccc} 
    \hline
	\hline   
     & &  & \textbf{ID 29} & &  \\
     \hline
     Parameter & Unit & Part 1 & Part 2 & Part 3 & Adopted  \\
                  &      & Sectors 17, 18 & Sector 24 & Sectors 57, 58 & solution \\
		\hline
    Orbital Period ($P_{\rm orb}$) & [days] & 1.69051 $\pm$ 0.00003 &  1.69049 $\pm$ 0.00007 
    & 1.69051 $\pm$ 0.00001 & 1.6905 $\pm$ 0.00008 \\
    Time of periastron ($T_{\rm p}$) & [days] & 2458766.44754 $\pm$ 0.00046 & 
    2458957.47509 $\pm$ 0.00054 & 2458855.20011 $\pm$ 0.00362 & \\
    Orbital inclination ($i$)        & [deg] & 70.902 $\pm$ 5.641 & 69.659 $\pm$ 4.319 & 
    81.393 $\pm$ 0.302 & 73.985 $\pm$ 7.111 \\
    \noalign{\smallskip}
    Fractional radius of star A ($r_{\rm A}$) &  & $0.2597^{+0.0556}_{-0.0043}$ & 
    $0.2593^{+0.0015}_{-0.0013}$ & $0.2711^{+0.0074}_{-0.0068}$ & $0.2634^{+0.0049}_{-0.0049}$ \\
    \noalign{\smallskip}
    Fractional radius of star B ($r_{\rm B}$) &  & $0.2252^{+0.0039}_{-0.0041}$ & 
    $0.2356^{+0.0015}_{-0.0016}$ & $0.2562^{+0.0054}_{-0.0047}$ & $0.2391^{+0.0044}_{-0.0048}$ \\
    \noalign{\smallskip}
    Light ratio ($l_{\rm B}/l_{\rm A}$) & & $0.5319^{+0.14503}_{-0.11617}$ & $0.6102^{+0.06415}_{-0.05960}$ & $0.6246^{+0.0394}_{-0.0358}$ & $0.5889^{+0.1731}_{-0.1468}$ \\
    \hline
    & &  & \textbf{ID 67} & &  \\
    \hline
    Orbital Period ($P_{\rm orb}$) & [days] & 24.49314 $\pm$ 0.00031 &  24.41614 $\pm$ 0.00255 & 
    24.49260 $\pm$ 0.0005 & 24.49257 $\pm$ 0.0005 \\
    Time of periastron ($T_{\rm p}$) & [days] & 2458767.08336 $\pm$ 0.00022 & 
    2458975.34104 $\pm$ 0.00024 & 2459857.01723 $\pm$ 0.00011 & \\
    Orbital inclination ($i$)        & [deg] & 87.497 $\pm$ 0.026 & 88.302 $\pm$ 0.016 & 
    88.454 $\pm$ 0.017 & 88.398 $\pm$ 0.026 \\
    \noalign{\smallskip}
    Orbital eccentricity($e$)        &      & $0.591^{+0.014}_{-0.007}$ & $0.595^{+0.003}_{-0.005}$ & 
    $0.595^{+0.010}_{-0.005}$ & $0.589^{+0.010}_{-0.005}$ \\
    \noalign{\smallskip}
    Periastron longitute omega ($w$)  & [deg] & $270.372^{+0.013}_{-0.007}$ & $269.403^{+0.011}_{-0.029}$ 
    & $269.597^{+0.013}_{-0.005}$ & $269.598^{+0.0106}_{-0.008}$ \\
    \noalign{\smallskip}
    Fractional radius of star A ($r_{\rm A}$) &  & $0.0488^{+0.0011}_{-0.0023}$ & 
    $0.0380^{+0.0002}_{-0.0003}$ & $0.0351^{+0.0049}_{-0.0049}$ & $0.0367^{+0.0005}_{-0.0006}$ \\
    \noalign{\smallskip}
    Fractional radius of star B ($r_{\rm B}$) &  & $0.0302^{+0.0003}_{-0.0005}$ & 
    $0.0275^{+0.0001}_{-0.0002}$ & $0.0259^{+0.0044}_{-0.0048}$ & $0.0272^{+0.0011}_{-0.0007}$ \\
    \noalign{\smallskip}
    Light ratio ($l_{\rm B}/l_{\rm A}$) & & $3.051^{+0.086}_{-0.156}$ & $0.135^{+0.001}_{-0.001}$ & $0.161^{+0.173}_{-0.147}$ & $0.168^{+0.007}_{-0.004}$ \\
    \hline
       &  &  & \textbf{ID 116} &  &  &   \\
       \hline
    Orbital Period ($P_{\rm orb}$) & [days] & 7.56667 $\pm$ 0.00004 &  7.56675 $\pm$ 0.00009 & 
    7.56662 $\pm$ 0.00013 & 7.56668 $\pm$ 0.00016 \\
    Time of periastron ($T_{\rm p}$) & [days] & 2458765.83110 $\pm$ 0.00015 & 
    2458962.56506 $\pm$ 0.00013 & 2459893.26611 $\pm$ 0.00015 & \\
    Orbital inclination ($i$)        & [deg] & 86.380 $\pm$ 0.180 & 86.041 $\pm$ 0.220 & 
    85.794 $\pm$ 0.217 & 86.071 $\pm$ 0.358 \\
    \noalign{\smallskip}
    Fractional radius of star A ($r_{\rm A}$) &  & $0.0743^{+0.0042}_{-0.0043}$ & 
    $0.0805^{+0.0015}_{-0.0013}$ & $0.0822^{+0.0021}_{-0.00195}$ & $0.0789^{+0.0049}_{-0.0049}$ \\
    \noalign{\smallskip}
    Fractional radius of star B ($r_{\rm B}$) &  & $0.0684^{+0.0039}_{-0.0041}$ & 
    $0.0770^{+0.0015}_{-0.0016}$ & $0.0763^{+0.0015}_{-0.0018}$ & $0.0739^{+0.0044}_{-0.0048}$ \\
    \noalign{\smallskip}
    Light ratio ($l_{\rm B}/l_{\rm A}$) & & $0.6939^{+0.14503}_{-0.11617}$ & $0.7506^{+0.06415}_{-0.05960}$ & $0.69461^{+0.06950}_{-0.06711}$ & $0.713^{+0.1731}_{-0.1468}$ \\
        \hline
        \hline
	\end{tabular}
\end{table*}

\begin{table*}
	\caption{Spectral properties (effective temperature - $T_{\rm eff}$,
 metallicity - $[Fe/H]$, surface gravity - $logg$) and MESA model results ($T_{\rm eff}$, $[Fe/H]$, $logg$ and stellar mass - $M$) of member binary 
        stars (system - A+B and components - A and B) with their IDs of UPK220 are presented. The spectral properties are derived by GSP-Phot and SED methods. Initial Helium and metallicity abundances are taken $Y_{\rm int}=0.2551$ and $Z_{\rm int}=0.0041$, respectively, for the MESA models of the binary member variable stars.}
	\label{tab:magbin}
    \begin{tabular}{rlllccl} 
		\hline
        ID & $T_{\rm eff}$ & [Fe/H] & logg & M & System & Method \\
           &  (K) & (dex) & (dex)& (M/${M_\odot}$) & \\
		\hline
      29 & $9565^{+178}_{-95}$ & $+0.31^{+0.09}_{-0.12}$ & $3.76^{+0.03}_{-0.02}$ & - & A+B & GSP-Phot \\
    \noalign{\smallskip}
       & $9542^{+196}_{-157}$ & $-0.55^{+0.01}_{-0.01}$ & $3.96^{+0.11}_{-0.12}$ & - & A+B & SED \\
       \noalign{\smallskip}
     &   11481 & $-0.56$ & 4.26 & 3.2 & A & MESA \\
     &    8912 & $-0.56$ & 4.06 & 2.0 & B & MESA \\
    \noalign{\smallskip}
        67 & $7330^{+92}_{-71}$ & $-1.08^{+0.05}_{-0.05}$ & $3.95^{+0.01}_{-0.01}$ & - & A+B & GSP-Phot \\
    \noalign{\smallskip}
       & $7522^{+86}_{-59}$ & $-0.61^{+0.06}_{-0.08}$ & $3.70^{+0.17}_{-0.15}$ & - & A+B & SED \\
    \noalign{\smallskip}
      &  11220 & $-0.56$ & 4.27 & 3.1 & A & MESA \\
      &   7244 & $-0.56$ & 4.09 & 1.5 & B & MESA \\
    \noalign{\smallskip}
    116 & $9216^{+48}_{-30}$ & $-0.76^{+0.05}_{-0.02}$ & $3.78^{+0.01}_{-0.01}$ & - & A+B & GSP-Phot \\
        \noalign{\smallskip}
        &$4600^{+100}_{-100}$ & $-1.00^{+0.25}_{-0.25}$ & $3.50^{+0.25}_{-0.25}$ & - & A & SED \\
    \noalign{\smallskip}
        &$5000^{+100}_{-100}$ & - & - & - & B & SED \\
        \noalign{\smallskip}
        & 6025 & $-0.56$ & 4.43 & 1.2 &  A & MESA \\
        & 4853 & $-0.56$ & 4.63 & 0.8 &  B & MESA \\
        \noalign{\smallskip}
    \hline
	\end{tabular}
\end{table*}

\begin{table*}
	\caption{Same as in Table~\ref{tab:magbin} but for single variable 
        stars. For ID16, spectral properties are derived using GSP-Spec 
        method.}
	\label{tab:specphoto_variable_table}
	\begin{tabular}{llllccl} 
		\hline
        ID & $T_{\rm eff}$ & [Fe/H] & logg & M & R & Method \\
           &  (K) & (dex) & (dex)& (M/${M_\odot}$) & (R/${R_\odot}$) & \\
		\hline
  \noalign{\smallskip}
      16 & $7550^{+250}_{-435}$ & - & $3.95^{+0.36}_{-0.15}$ & - & 7.5 & GSP-Spec \\
    \noalign{\smallskip}
         & $7511^{+68}_{-59}$ & $-0.58^{+0.07}_{-0.09}$ & $3.93^{+0.14}_{-0.12}$ & - & - &  SED \\
    \noalign{\smallskip}
   &  7638 & $-0.56$ & 4.41 & 1.3 & 1.2 & MESA \\
    \noalign{\smallskip}
    42 & $7997^{+2}_{-6}$ & $-0.16^{+0.09}_{-0.15}$ & $4.07^{+0.02}_{-0.01}$ & - & 1.9 &GSP-Phot \\
    \noalign{\smallskip}
       & $7470^{+80}_{-111}$ & $-0.37^{+0.07}_{-0.11}$ & $4.03^{+0.11}_{-0.12}$ & - & - &SED \\
    \noalign{\smallskip}
    &  8390 & $-0.56$ & 4.44 & 1.4 &  1.2 &MESA \\
    \noalign{\smallskip}
    49 & $8577^{+74}_{-82}$ & $-0.47^{+0.09}_{-0.06}$ & $4.20^{+0.04}_{-0.07}$ & - & 1.7 &GSP-Phot \\
    \noalign{\smallskip}
       & $8353^{+71}_{-89}$ & $-0.53^{+0.06}_{-0.09}$ & $4.60^{+0.12}_{-0.11}$ & - &  - &SED \\
    \noalign{\smallskip}
    &  8491 & $-0.56$ & 4.45 & 1.5 & 1.2 &MESA \\
    \noalign{\smallskip}
    138 &  $4419^{+81}_{-100}$ & $-0.64^{+0.21}_{-0.13}$ & $4.71^{+0.02}_{-0.02}$ & - & 0.7 &GSP-Phot \\
    \noalign{\smallskip}
        & $4939^{+118}_{-90}$ & $-0.58^{+0.10}_{-0.14}$ & $4.69^{+0.11}_{-0.14}$ & - & - &SED \\
        \noalign{\smallskip}
        & 4508 & $-0.56$ & 4.73 & 0.6 & 0.6 &MESA \\
    \noalign{\smallskip}
    147 &  $5412^{+51}_{-41}$ & $-0.57^{+0.07}_{-0.09}$ & $4.59^{+0.03}_{-0.02}$ & - & 1.0 &GSP-Phot \\
    \noalign{\smallskip}
        & $5300^{+108}_{-85}$ & $-0.61^{+0.06}_{-0.09}$ & $4.22^{+0.15}_{-0.13}$ & - & - &SED \\
    \noalign{\smallskip}
     & 5598 & $-0.56$ & 4.64 & 0.8 & 0.7 &MESA \\
    \noalign{\smallskip}
    \hline
	\end{tabular}
\end{table*}

\begin{table*}
	\caption{Lomb-Scargle periodogram analysis of {\it TESS} observations of ID67 
 and ID 116. Derived 
      frequencies of pulsation (in c/d unit) with their amplitudes (in mag unit), 
        phases (in rad/2$\pi$ unit)
        and estimated periods (in days unit) from the residuals are 
 presented with their uncertainties. S1718, S24, and S5758 refer to 
 sectors 17 and 
        18, 24, and 57 and 58 of {\it TESS} observations, respectively.}
	\label{tab:tableper3}
	\begin{tabular}{llcccr}
		\hline 
        \hline 
        ID & Frequency (c/d) & Amplitude (mag) & Phase (rad/2$\pi$) & Period (day) &  Sectors \\
        \hline
        67 & 0.8496901 $\pm$ 0.0021476 ($f$) & 0.0024 $\pm$ 4.66e-05 & 0.0844 $\pm$ 0.0031 & 1.177 $\pm$ 0.006 &  S1718 \\
         & 0.8434911 $\pm$ 0.0005211 ($f$) & 0.0013 $\pm$ 3.18e-05 & 0.3893 $\pm$ 0.0041 & 1.186 $\pm$ 0.002 &  S24 \\
        & 0.8449160 $\pm$ 0.0001094 ($f$) & 0.0023 $\pm$ 2.55e-05 & 0.0256 $\pm$ 0.0018 & 1.184 $\pm$ 0.001 &  S5758 \\
        \hline
        116 & 0.1842825 $\pm$ 0.0024367 ($f$) & 0.0002 $\pm$ 1.88e-05 & 0.0997 $\pm$ 0.0139 & 5.426 $\pm$ 0.139 & S5758 \\
        & 0.3415368 $\pm$ 0.0025788 ($f/2$) & 0.0002 $\pm$ 1.88e-05 & 0.8737 $\pm$ 0.0151 & 2.928 $\pm$ 0.044 & S5758 \\
        \hline
        \hline
	\end{tabular}
\end{table*}

\begin{table*}
	\caption{Lomb-Scargle periodogram analysis of {\it TESS} 
     observations in all sectors of ID 49. Derived frequencies of 
     pulsation (in c/d unit) with their amplitudes (in mag unit) and 
     phases (in rad/2$\pi$ unit) are presented.}
	\label{tab:gammador}
    \begin{tabular}{lccc}
     ID & Frequency (c/d) & Amplitude (mag) & Phase (rad/2$\pi$) \\
\hline
f1&     0.0202850 &  0.0022 &  0.2341 \\
f2&     1.5019749 &  0.0018 &  0.5717 \\
f3&     0.0511536 &  0.0021 &  0.4037 \\
f4&     0.0679108 &  0.0017 &  0.4499 \\
f5&     0.1331757 &  0.0021 &  0.5337 \\
f6&     1.3529241 &  0.0012 &  0.7303 \\
f7&     0.1173004 &  0.0018 &  0.9738 \\
f8&     0.1737458 &  0.0016 &  0.3178 \\
f9&     0.1922669 &  0.0011 &  0.2162 \\
f10&    0.0987793 &  0.0007 &  0.1113 \\
f11&    1.3767368 &  0.0006 &  0.4531 \\
f12&    0.2654694 &  0.0007 &  0.2894 \\
f13&    0.2196076 &  0.0008 &  0.8653 \\
f14&    0.1552246 &  0.0007 &  0.2997 \\
f15&    0.0837861 &  0.0008 &  0.1004 \\
f16&    0.2407746 &  0.0005 &  0.2581 \\
f17&    0.0335144 &  0.0008 &  0.3487 \\
f18&    1.5240239 &  0.0003 &  0.7271 \\
f19&    1.2250401 &  0.0003 &  0.3399 \\
f20&    1.6289769 &  0.0003 &  0.7256 \\
f21&    0.3280885 &  0.0003 &  0.9015 \\
\hline
    \end{tabular}
    \end{table*}

\begin{table*}
\tiny
\caption{Fundamental parameters comparison with the literature for UPK~220.}
\setlength{\tabcolsep}{0.05cm}
\begin{tabular}{cccccccll}
\hline 
E(B-V) &({$V_{\rm 0}$}-{$M_{\rm V}$})& d~(kpc) & [Fe/H] &log(Age) & Age~(Myr)& Isochrone & Photometry & Ref.\\
\hline 
1.3 & 9.60 & 0.967 & -0.56 & 8.15 & 140.0 & MIST & GaiaDR3,~G, {G${_{\rm BP}}$}, {G${_{\rm RP}}$} & This paper \\
 - & - & 0.967 $\pm$ 0.28 & 0.127
 & 8.75 & 562.34 & PARSEC & GaiaDR2~Gaia~G, {G${_{\rm BP}}$}, {G${_{\rm RP}}$} & \citet{2019JKAS...52..145S} \\
0.70 & 9.98 & 0.99 & 0.0 & 8.04 & 109.65 & PARSEC & GaiaDR2,~Gaia~G, {G${_{\rm BP}}$}, {G${_{\rm RP}}$} & \citet{2020AA.640A..1C} \\
 0.880 $\pm$ 0.071 & - & 0.950 $\pm$ 0.12 & 0.029$\pm$ 0.129 & 7.295 $\pm$ 0.329 & 19.72 & PARSEC & GaiaDR2,~Gaia~G, {G${_{\rm BP}}$}, {G${_{\rm RP}}$} & \cite{2021MNRAS.504..356D}\\
-  & - & 0.964 & 0.0 & 8.08 & 120.23 & PARSEC & Gaia EDR3,~Gaia~G, {G${_{\rm BP}}$}, {G${_{\rm RP}}$} & \citet{tarricq2022structural} \\
0.90 $\pm$ 0.052 & - & 0.944 $\pm$ 0.009 & 0.025$\pm$ 0.097 & 7.477$\pm$ 0.290 & 29.99 & PARSEC &Gaia EDR3,~ Gaia~G, {G${_{\rm BP}}$}, {G${_{\rm RP}}$} & \citet{almeida2023revisiting} \\
0.98& 10.31 & 1.153 & 0.75 & 8.01 & 102.33 & PARSEC & GaiaDR3,~Gaia~G, {G${_{\rm BP}}$}, {G${_{\rm RP}}$}; 2MASS~J,H & \citet{cavallo2023parameter} \\
0.70 & 9.98 & 0.997 & 0.301 & 7.72 & 52.48 & PARSEC & GaiaDR3~G, {G${_{\rm BP}}$}, {G${_{\rm RP}}$} & \citet{2024AA...689A..18A} \\

\hline
\label{tab:literature}
\end{tabular}
\end{table*}

\vspace{5mm}
\facilities{Gaia(DPAC), TESS(SPOC)}

\software{astropy \citep{2018AJ....156..123T},  
          matplotlib \citep{Hunter:2007}, 
          numpy \citep{harris2020array},
          astroquery \citep{2019AJ....157...98G},
          Lightkurve \citep{Lightkurve},
          MESA \citep{2011ApJS..192....3P, 2015ApJS..220...15P, 2018ApJS..234...34P, 
          2019ApJS..243...10P},
          TESScut \citep{2019ascl.soft05007B},
          ARIADNE \citep{10.1093/mnras/stac956},
          JKTEBOP \citep{2004MNRAS.351.1277S}}

\bibliography{upk220}{}
\bibliographystyle{aasjournal}

\appendix

\section{SED Photometric Data}

The SED fitting performed using Bayesian Model Averaging to obtain the best-fitted
model parameters. To fit each SED of the variable member stars, 
we collected \textit{2MASS} (J,H and $K_s$), \textit{Gaia DR3} (G, ${G_{\rm BP}}$, and ${G_{\rm RP}}$), \textit{Johnson} (U, B, V, R, and
I), \textit{PS1} (g,i, r, y, and z), \textit{SDSS} (u, g, r, and i), \textit{TESS} (Tmag), 
and \textit{WISE} (W1, W2, W3 and W4) photometric data. Unfortunately, we were unable to obtain all the photometric data for each star.
The data used in the SED fitting  are listed in 
Table~\ref{tab:variable_sedphoto}.

\begin{longrotatetable}
\begin{deluxetable*}{lccccccccc}
\tablecaption{The collected photometric data with their errors 
from various catalogs of the variable member stars for SED fitting.}
\label{tab:variable_sedphoto}
\tablewidth{700pt}
\tabletypesize{\scriptsize}
\tablehead{
\colhead{Survey/Filter} & \colhead{ID16} & 
\colhead{ID29} & \colhead{ID42} & 
\colhead{ID49} & \colhead{ID67} & 
\colhead{ID116} & \colhead{ID138} & 
\colhead{ID147} & \colhead{ID167} \\ 
} 
\startdata
     2MASS-H & 9.644 $\pm$ 0.029 & 10.322 $\pm$ 0.032 & 12.359 $\pm$ 0.031 & 12.177 $\pm$ 0.030 & 11.327 $\pm$ 0.035 & 10.838 $\pm$ 0.028 & 14.433 $\pm$ 0.048 & 13.784 $\pm$ 0.046 & 9.558 $\pm$ 0.026 \\
     2MASS-J & 9.959 $\pm$ 0.023 & 10.595 $\pm$ 0.024 & 12.827 $\pm$ 0.025 & 12.513 $\pm$ 0.023 & 11.670 $\pm$ 0.026 & 11.055 $\pm$ 0.022 & 15.221 $\pm$ 0.052 & 14.399 $\pm$ 0.032 & 10.011 $\pm$ 0.023 \\
     2MASS-{$K_{\rm s}$} & 9.494 $\pm$ 0.024 & 10.126 $\pm$ 0.021 & 12.255 $\pm$ 0.026 & 12.062 $\pm$ 0.027 & 11.164 $\pm$ 0.024 & 10.691 $\pm$ 0.023 & 14.138 $\pm$ 0.079 & 13.600 $\pm$ 0.044 & 9.337 $\pm$ 0.022 \\
     \hline
     GaiaDR3-G & 11.976 $\pm$ 0.002 & 12.636 $\pm$ 0.001 & 15.226 $\pm$ 0.003 &14.164 $\pm$ 0.001 & 14.148 $\pm$ 0.001 & 12.619 $\pm$ 0.002 & 18.043 $\pm$ 0.002 & 16.803 $\pm$ 0.001 & 12.667 $\pm$ 0.001 \\
     GaiaDR3-${G_{\rm BP}}$ & 12.704 $\pm$ 0.001 & 13.418 $\pm$ 0.005 & 16.203 $\pm$ 0.004 & 14.737 $\pm$ 0.002 & 14.791 $\pm$ 0.019 & 13.153 $\pm$ 0.001 & 19.313 $\pm$ 0.041 & 17.741 $\pm$ 0.006 & 13.823 $\pm$ 0.004 \\
     GaiaDR3-${G_{\rm RP}}$ & 11.134 $\pm$ 0.001 & 11.760 $\pm$ 0.003 & 14.238 $\pm$ 0.001 & 13.442 $\pm$ 0.001 & 13.089 $\pm$ 0.027 & 11.928 $\pm$ 0.001 & 16.921 $\pm$ 0.005 & 15.825 $\pm$ 0.003 & 11.593 $\pm$ 0.003 \\
     \hline
     Johnson-B & - & 15.539 $\pm$ 0.210 & - & - & - & 17.143 $\pm$ 0.305 & - & - & -  \\
     Johnson-V & - & 14.460 $\pm$ 0.098 & - & - & - & 16.329 $\pm$ 0.021 & - & - & -  \\
     \hline
     PS1-g &- & - & 16.554 $\pm$ 0.004 & 14.859 $\pm$ 0.002 & 14.911 $\pm$ 0.016 & 12.996 & 19.907 $\pm$ 0.015 & 18.091 $\pm$ 0.007 & 14.249 $\pm$ 0.003 \\
     PS1-i &- & - & 14.677 $\pm$ 0.003 & 13.834 $\pm$ 0.005 & 13.142  & 12.307 & 17.395 $\pm$ 0.004 & 16.256 $\pm$ 0.003 & 12.024 \\
     PS1-r &- & - & 15.359 $\pm$ 0.003 & 14.211 $\pm$ 0.002 & 14.091 $\pm$ 0.036 & 12.553 & 18.317 $\pm$ 0.005 & 16.898 $\pm$ 0.008 & 12.753 \\
     PS1-y &- & - & 14.005 $\pm$ 0.004 & 13.467 $\pm$ 0.003 & 13.293 $\pm$ 0.027 & 11.951 & 16.525 $\pm$ 0.006 & 15.607 $\pm$ 0.004 & 11.329 \\
     PS1-z &- & - & 14.286 $\pm$ 0.002 & 13.632 $\pm$ 0.002 & 12.869  & 12.093 & 16.882 $\pm$ 0.003 & 15.888 $\pm$ 0.003 & 11.475 \\
     \hline
     SDSS-g &- & 14.958 $\pm$ 0.156 & - & - & 16.932 $\pm$ 0.289 & 17.371 $\pm$ 0.083 & - & - & - \\
     SDSS-r &- & 14.068 $\pm$ 0.081 & - & - & 15.606 $\pm$ 0.022 & 15.971 & - & - & - \\
     SDSS-i &- & 13.571 $\pm$ 0.079 & - & - & 15.282 $\pm$ 0.033 & 15.123 & - & - & - \\
     \hline
     TESS & 11.212 $\pm$ 0.051 & 11.788 $\pm$ 0.023 & 14.219 $\pm$ 0.021 & 13.503 $\pm$ 0.045 & 13.335 $\pm$ 0.007 & 12.012 $\pm$ 0.037 & 16.830 $\pm$ 0.151 & 15.821 $\pm$ 0.0381 & 11.527 $\pm$ 0.051 \\
     \hline
     WISE-W1 & 9.253 $\pm$ 0.021 & 10.020 $\pm$ 0.022 & 12.064 $\pm$ 0.024 & - & - & 10.577 $\pm$ 0.023 & - & - & 9.168 $\pm$ 0.023 \\
     WISE-W2 & 9.263 $\pm$ 0.019  & 9.998 $\pm$ 0.019 & 12.039 $\pm$ 0.022 & - & - & 10.575 $\pm$ 0.021 & - & - & 9.127 $\pm$ 0.021 \\
     \hline
     \hline
\enddata
\end{deluxetable*}
\end{longrotatetable}

\end{document}